\pdfoutput=1
\documentclass[aip,graphicx]{revtex4-1}
\usepackage{graphicx}
\usepackage{dcolumn}
\usepackage{bm}
\usepackage{amsmath}
\usepackage{mathrsfs}
\usepackage{subfigure}
\usepackage{float}
\usepackage{natbib}
\usepackage{threeparttable}
\usepackage{hyperref}
\usepackage{color}

\draft 

\begin{document}

\title{A hybrid numerical algorithm based on the stochastic particle Shakhov and DSMC method} 



\author{Hao Jin}
\email[]{jinhao@mail.nwpu.edu.cn}
\affiliation{School of Aeronautics, Northwestern Polytechnical University, Xi'an, Shaanxi 710072, China}
\author{Sha Liu}
\email[Corresponding author: ]{shaliu@nwpu.edu.cn}
\affiliation{School of Aeronautics, Northwestern Polytechnical University, Xi'an, Shaanxi 710072, China}
\affiliation{Institute of Extreme Mechanics, Northwestern Polytechnical University, Xi'an, Shaanxi 710072, China}
\affiliation{National Key Laboratory of Aircraft Configuration Design, Northwestern Polytechnical University, Xi'an, Shaanxi 710072, China}
\author{Sirui Yang}
\affiliation{School of Aeronautics, Northwestern Polytechnical University, Xi'an, Shaanxi 710072, China}
\author{Junzhe Cao}
\affiliation{School of Aeronautics, Northwestern Polytechnical University, Xi'an, Shaanxi 710072, China}
\author{Congshan Zhuo}
\affiliation{School of Aeronautics, Northwestern Polytechnical University, Xi'an, Shaanxi 710072, China}
\affiliation{Institute of Extreme Mechanics, Northwestern Polytechnical University, Xi'an, Shaanxi 710072, China}
\affiliation{National Key Laboratory of Aircraft Configuration Design, Northwestern Polytechnical University, Xi'an, Shaanxi 710072, China}
\author{Chengwen Zhong}
\affiliation{School of Aeronautics, Northwestern Polytechnical University, Xi'an, Shaanxi 710072, China}
\affiliation{Institute of Extreme Mechanics, Northwestern Polytechnical University, Xi'an, Shaanxi 710072, China}
\affiliation{National Key Laboratory of Aircraft Configuration Design, Northwestern Polytechnical University, Xi'an, Shaanxi 710072, China}


\date{\today}

\begin{abstract}
    The Direct Simulation Monte Carlo (DSMC) method provides a particle-based numerical solution to the Boltzmann equation, and is widely employed for simulating rarefied nonequilibrium gas flows. With advances in aerospace engineering and micro/nano-scale technologies, gas flows increasingly exhibit the coexistence of rarefied and continuum/near-continuum regimes, which calls for larger time steps and coarser spatial grids for efficient and accurate numerical simulation. However, the mesh sizes and time steps in DSMC are constrained by the single-scale nature of the Boltzmann equation and the explicit treatment of collision term following operator splitting. To overcome the resulting computational inefficiency, the Time-Relaxed Monte Carlo (TRMC) method introduces a suitable time discretization of the Boltzmann equation, allowing for significantly larger time steps. Besides, domain decomposition methods leverage the complementary strengths of continuum and particle-based approaches, facilitating the efficient simulation of multi-scale gas flows. However, in TRMC method, the physically accurate high-order terms are truncated and approximated through convergence to a local Maxwellian distribution. Meanwhile, the continuum breakdown criteria employed in hybrid methods are either empirical or semi-empirical. Therefore, it is essential to develop a more concise and rational method for extending the applicability of DSMC to a broader range of flow regimes. Recently, a timescale-based decomposition of the Boltzmann equation has been proposed to enable a more rational coupling between DSMC and Navier-Stokes. Inspired by this strategy, a novel hybrid particle method is proposed to couple the stochastic particle Shakhov with DSMC, in which the collision operator is decomposed into two sub-steps based on local observation timescale and the relaxation time. The validity and accuracy of the proposed method are demonstrated through a series of benchmark cases, including 1-D sod shock tube, 2-D hypersonic flow around cylinder and jet expansion into the vacuum, 3-D hypersonic flows around sphere and X-38 like vehicle in near-continuum flow regimes.
\end{abstract}


\maketitle 
\section{Introduction}\label{Introduction}
	Multi-scale gas flows are ubiquitous in aerospace engineering applications\cite{gnoffo1999planetary, bertin2006critical}, such as hypersonic flight in near-space environments, spacecraft re-entry, and satellite attitude control. These flows are characterized by the presence of different flow regimes, from the continuum one to the rarefied one. Accurately and efficiently simulating these multi-scale gas flows is essential for the design and optimization of aerospace vehicles and systems\cite{schouler2020survey}. Traditional computational fluid dynamics (CFD) methods, based on the Navier-Stokes (N-S) equations\cite{blazek2015computational}, are derived under the continuum assumption and become invalid in rarefied regimes\cite{ivanov1998computational}. In contrast, the Direct Simulation Monte Carlo (DSMC) method\cite{bird1994molecular}, a particle-based approach that directly models the physical process between gas particles, is widely used to simulate rarefied gas flows. However, DSMC suffers from prohibitively high computational cost in the continuum and near-continuum regimes, as its grid size and time step must remain smaller than the local mean free path and mean collision time, respectively.  As a result, neither of these methods alone is sufficient for efficiently and accurately simulating multi-scale gas flows.

    To simulate multi-scale gas flows, a straightforward strategy is to employ the DSMC method in rarefied regions and the N-S method in continuum regions, resulting in hybrid approaches such as the modular particle-continuum (MPC) method\cite{schwartzentruber2007modular}. In hybrid N-S/DSMC methods\cite{schwartzentruber2007modular, wadsworth1992two, sun2004hybrid, burt2009hybrid, schwartzentruber2015progress, xu2018parallelized, vasileiadis2024hybriddcfoam}, two critical issues must be considered: (1) Determination of the interface position between the DSMC and N-S regions. To balance efficiency and accuracy, numerous continuum breakdown criteria have been proposed, such as $P$ parameter\cite{bird1970breakdown}, Knudsen number\cite{boyd1995predicting, wang2002continuum, lian2011improved}, and $B$ parameter\cite{garcia1999adaptive}. Among them, the $\mathrm{Kn}_Q$ number\cite{wang2002continuum}, defined as the ratio of local mean free path to the gradients of macroscopic variables, is a widely used breakdown criterion for determining the interface. However, these criteria are either empirical or semi-empirical, and there is no universal breakdown criterion applicable to all flow conditions. (2) Properly information exchanging at the interface. To ensure proper information exchange between continuum and particle solvers, either flux-based data at the interface\cite{hash1996assessment} or macroscopic variables within buffer zones\cite{sun2004hybrid} are typically employed. This process is inherently challenging, as the DSMC solver operates on microscopic molecular information, whereas the N-S solver relies on macroscopic flow variables. For information transfer from N-S to DSMC solver, molecular properties should be reconstructed from macroscopic flow variables using approximate expressions such as the Maxwellian distribution\cite{bird1994molecular} or Chapman-Enskog (C-E) expansion\cite{garcia1998generation}. However, these approximations are only accurate in near-continuum regions. Conservely, when transferring information from DSMC to N-S solver, molecular stochastic fluctuations can destabilize the N-S computations, necessitating the use of buffer zones. Therefore, improving physical accuracy and reducing statistical fluctuations remain persistent challenges in hybrid DSMC/N-S methods.

    To reduce statistical fluctuations during information exchange, the N-S method can be replaced with a particle-based method. Several hybrid particle methods\cite{burt2009hybrid, kumar2013development, jiang2016improved, malaikannan2017hybrid, jun2018assessmentLD, pfeiffer2019evaluation, zhang2019particle, fei2021hybrid, kim2024evaluation, vasileiadis2025unigasfoam} have been proposed, primarily differing in their treatments of the continuum regime. These approaches can be broadly categorized into continuum (single-scale) particle methods and multi-scale particle methods. For continuum particle methods, representative approaches include equilibrium particle simulation method (EPSM)\cite{pullin1980direct}, Collision Limiter (CL) method\cite{bartel1994dsmc}, and Low Diffusion (LD) method\cite{burt2008low}. These methods enforces the molecules following the ``equilibrium" distribution in the continuum limit using a modified DSMC collision term, and are coupled with DSMC method to simulate multi-scale gas flows\cite{jiang2016improved, malaikannan2017hybrid, jun2018assessmentLD}. In contrast, the multi-scale methods are based on the Boltzmann model equations rather than DSMC's binary collision model. These include methods based on Fokker-Planck (FP)-type models\cite{gorji2011fokker, gorji2015fokker, jun2018assessment, kim2024evaluation} and Bhatnagar-Gross-Krook (BGK)-type models\cite{bhatnagar1954model, pfeiffer2018particle}. The Stochastic Particle (SP) methods\cite{macrossan2001nu}, based on BGK-type relaxation processes, allow for larger time steps and coarser spatial grids than DSMC while maintaining comparable accuracy in near-continuum regimes\cite{pfeiffer2018particle, pfeiffer2025crank}. However, these methods exhibit additional numerical viscosity in the continuum regime, since the molecular free transport shares the same physical nature as first-order upwind scheme in CFD. To address this issue, the Unified Stochastic Particle (USP) methods\cite{fei2020unified} incorporates the particle transport process into the collision term, and anti-dissipation (anti-viscosity) is added physically by a Grad-like distribution. However, a slight discrepancy arises between the model Boltzmann equation and the original one when the observation time (or numerical iteration step) is shorter than the mean collision time. By combining the computational efficiency of SP and USP methods with the accuracy of DSMC, these methods are naturally coupled to simulate multi-scale gas flows\cite{kumar2013development, fei2021hybrid, fei2022unified}, and no instabilities occur unlike in the hybrid N-S/DSMC methods. However, the criteria used in hybrid particle-particle methods above are also empirical or semi-empirical, and there is no universal breakdown criterion.

	Different from the domain decomposition methods above, an alternative multi-scale strategy achieves coupling at the algorithmic level. A notable example is the Time Relaxed Monte Carlo (TRMC) method\cite{pareschi2001time, fei2023time}, which introduces a novel approach by applying a suitable temporal discretization of the Boltzmann equation. Starting from the Wild sum expansion\cite{carlen2000central}, TRMC method constructs a family of Monte Carlo schemes by approximating higher-order terms with local Maxwellian distributions, achieving both high-order temporal accuracy and asymptotic preservation. In contrast to DSMC, TRMC eliminates strict time-step constraints, enabling high-fidelity simulations with significantly larger time steps. Moreover, it naturally satisfies the asymptotic preserving property: as the Knudsen number tends to zero, the solution automatically relaxes to the local Maxwellian distribution, seamlessly recovering the hydrodynamic limit. However, the selection of coefficients for high-order schemes remains empirical, and the recursive structure may reduce computational efficiency in practical implementations.
	
    In addition to the TRMC method, continuum and particle solvers can also be hybridized at the algorithmic level within a computational cell. A notable example is the Unified Gas-Kinetic Wave-Particle particle (UGKWP) method\cite{liu2020unified}, which has been widely applied to simulate multi-scale gas flows\cite{chen2020three, wei2024unified, long2024nonequilibrium}. Based on the analytical solution of Boltzmann BGK-type model, the UGKWP method couples stochastic particles (particle part) with macroscopic methods (wave part, i.e., the gas-kinetic scheme (GKS)\cite{xu2001gas}) at the algorithmic level. Furthermore, a Quantified Model Competition (QMC) mechanism\cite{liu2020simplified} is extracted from the analytical (time integral) solution of BGK-type model equation, for which the physical interpretation is that a fraction ${{e}^{-\Delta t/\tau }}$ of molecules are free transport ones while remaining fraction $1-{{e}^{-\Delta t/\tau }}$ of molecules follows the N-S mechanism; the latter can be eliminated and represented by macroscopic variables instead. This directly and explicitly provides a hybrid algorithm strategy, such as Simplified Unified Wave-Particle method (SUWP)\cite{liu2020simplified} and Simple Hydrodynamic-Particle Method (SHPM)\cite{liu2022simple}. However, since these methods are still based on the BGK-type models, their accuracy in simulating rarefied gas flows remains inferior to that of the DSMC method.

	To achieve a more physically consistent and rational coupling between DSMC and continuum methods (e.g., SP particle or N-S), multi-scale mechanisms can be decomposed according to their characteristic temporal scales. In Ref.\cite{liu2024multi}, the time integral solution of BGK-type models is used as a test function to perform a temporal decomposition of the Boltzmann equation. This decomposition reveals two distinct characteristic timescales: one associated with the non-equilibrium collision process described by the Boltzmann equation, and the other with the near-equilibrium relaxation captured by BGK-type models. Leveraging this mechanism, the DSMC method is incorporated to improve the modeling accuracy of SUWP method. In this work, we employ the above mechanism to design a coupling strategy and the corresponding algorithm between DSMC and SP Shakhov methods at the temporal hierarchy, thereby addressing the persistent reliance on empirical criteria in existing hybrid methods.

	The remainder of this paper is organized as follows. Section \ref{The kinetic theory} provides a brief overview of the Boltzmann equation, with a focus on the DSMC and stochastic particle Shakhov methods. Section \ref{hybrid method} introduces a novel hybrid algorithm for particle-based methods, constructed based on the local observation timescale and relaxation time. Building on this framework, a hybrid method that couples the stochastic particle Shakhov model with the DSMC method is proposed. Section \ref{Number simulations} presents  numerical simulations of several typical multi-scale gas flows to validate the accuracy of the proposed hybrid particle-particle method. Finally, concluding remarks are given in Section \ref{sec:Conclusion}.

\section{The Boltzmann equation and particle methods}\label{The kinetic theory}
	In gas kinetic theory\cite{chapman1990mathematical}, the Boltzmann equation serves as the foundational governing equation for the evolution of the particle distribution function. In the absence of external forces, it takes the form
	\begin{equation}
		\frac{\partial f}{\partial t}+\mathbf{c}\cdot \frac{\partial f}{\partial \mathbf{r}}={{J}_\mathrm{Boltzmann}}, \label{Boltzmann equation}
	\end{equation}
	where $f(\mathbf{c}, \mathbf{r}, t)$ denotes the gas particle distribution function in phase space, with $\mathbf{c}$ representing particle velocity and $\mathbf{r}$ describing particle position.  The term ${J}_\mathrm{Boltzmann}$ denotes the Boltzmann collision integral, defined as
	\begin{equation}
		{J}_\mathrm{Boltzmann}=\int_{-\infty}^{+\infty} \int_0^{4 \pi}\left(f^{\prime} f_1^{\prime}-f f_1\right) {\sigma}{{c}_{r}} d \Omega d \mathbf{c}_1,
	\end{equation}
	where $f^{\prime}$ and $f$ are the post-collision and pre-collision distribution functions, respectively, $c_r$ is the relative velocity between two colliding particles, and $\sigma$ and $\Omega$ represent the differential collision cross section and solid angel. Due to the nonlinearity and high-dimensional nature of the Boltzmann collision integral $J_{\text {Boltzmann}}$ in phase space and time, analytical solutions to the full Boltzmann equation are generally intractable. To overcome this challenge, two primary particle-based numerical approaches are commonly employed. A famous one is DSMC method\cite{stefanov2019basic}, which emulates the physics of the Boltzmann equation by tracking the free transport and collisions of a large number of simulated (model) particles. Another approach, called the stochastic particle method, adopts the simplified models, such as BGK-type\cite{bhatnagar1954model, pfeiffer2018particle} and FP-type\cite{gorji2011fokker, gorji2015fokker,jun2018assessment, kim2024evaluation} models, which approximates the Boltzmann collision term as a relaxation process towards equilibrium or simple intermolecular drift and diffusion forces. The remainder of this section provides a brief overview of these two computational approaches.

    \subsection{Direct simulation Monte Carlo}~{}
	\label{DSMC}
	The DSMC method is a stochastic, particle-based numerical approach for solving the Boltzmann equation\cite{bird1994molecular, 2013The, bird1970direct}. Here, the velocity distribution function is approximated by a finite ensemble of simulated particles, which can be mathematically represented as a superposition of Dirac delta functions in phase space. Through an operator splitting scheme, the original Boltzmann equation Eq.(\ref{Boltzmann equation}) is decomposed into free-transport and collision steps, i.e.,
	\begin{equation}
		\begin{aligned}
		& \frac{\partial f}{\partial t}+\mathbf{c}\cdot \frac{\partial f}{\partial \mathbf{r}}=0 \quad \text { and } \\
		& \frac{\partial f}{\partial t}={J}_\mathrm{Boltzmann}.
		\end{aligned}
	\end{equation}
	Notice that the fundamental principle of the DSMC method lies in decoupling the free-transport and collision process of simulated particles, where each computational particle represents a statistical ensemble of real molecules. During each time step $\Delta t$, the particle evolution is solved through two sequential sub-steps: free-transport and collision processes. In the collision step, particles maintain their spatial positions while stochastically undergoing binary collisions with neighboring particles within the same computational cell, resulting exclusively in velocity modifications.
	
	The free-transport step is treated deterministically, and the particles can be accurately tracked by solving the following equation
	\begin{equation}
		{{\mathbf{r}}^{n+1}_{i}}={{\mathbf{r}}^{n}_{i}}+{{\mathbf{c}}_{i}}\Delta t, \label{free transport}
	\end{equation}
	where subscript ``$i$'' denotes the particle index, $\mathbf{r}_i$ and $\mathbf{c}_i$ represent the position and velocity of the $i^{th}$ particle, respectively. According to Eq.(\ref{free transport}), this particle is advected to its new position and subsequently mapped to the computational cell containing it.

	The collision step is executed within each computational cell, where particle pairs are stochastically sampled for collision events. As a core component of the DSMC method, the collision algorithm determines the post-collision velocities of selected particle pairs based on their pre-collision states. The most widely employed collision scheme is the ``No Time Counter (NTC)'' scheme\cite{bird1994molecular, 2013The} which estimates the number of candidate collision pairs $N_\mathrm{col}$ per time step by utilizing the maximum collision probability. Here, the number of candidate particle pairs $N_\mathrm{col}$ is computed as follows
	\begin{equation}
		{{N}_\mathrm{col}}=\frac{1}{2}N\left( N-1 \right){{F}_{N}}{{\left( {{\sigma }_{T}}{{c}_{r}} \right)}_{\max }}\Delta t/{{V}_{c}}, \label{Ncol_DSMC}
	\end{equation}
	where $N$ is the number of simulated particles in the computational cell, $F_N$ denotes the ratio of real molecules to simulated particles, $\sigma_T$ represents the total collision cross section, $c_r$ is the relative velocity between two colliding particles, and $V_c$ is the volume of the computational cell. Particle pairs are selected randomly within the computational cell, and the collision probability $P_c$ for each pair is computed as
	\begin{equation}
		P_c=\frac{{{\sigma }_{T}}{{c}_{r}}}{{{\left( {{\sigma }_{T}}{{c}_{r}} \right)}_{\max }}},
	\end{equation}
	where ${{\sigma }_{T}}{{c}_{r}}$ represents the product of the total collision cross-section and the relative velocity of the selected particle pair. ${\left({{\sigma }_{T}}{{c}_{r}} \right)}_{\max }$ is initially assigned a suitable reference value within the computational cell and is subsequently updated if ${{\sigma }_{T}}{{c}_{r}}$ for a chosen particle pair exceeds this threshold. Comparing the collision probability $P_c$ with a random number $\mathscr{R}$, where $\mathscr{R}$ is a random number in $[0, 1]$, this acceptance-rejection (AR) procedure determines whether the candidate particle pair actually collides. Notice that the NTC scheme typically requires $10\sim40$ simulated particles per computational cell to provide accurate results\cite{bird1994molecular}. When a collision occurs, the post-collision velocities of particle pair are updated as follows
	\begin{equation}
		\begin{aligned}
			& \mathbf{c}_{1}^{\prime}=\frac{1}{2}[(\mathbf{c}_1+\mathbf{c}_2)-|\mathbf{g}|\mathbf{\omega}], \\
			& \mathbf{c}_{2}^{\prime}=\frac{1}{2}[(\mathbf{c}_1+\mathbf{c}_2)+|\mathbf{g}|\mathbf{\omega}],
		\end{aligned}
	\end{equation}
	where $\mathbf{c}_1$ and $\mathbf{c}_2$ denote the pre-collision velocities of the two colliding particles, $\mathbf{c}_{1}^{\prime}$ and $\mathbf{c}_{2}^{\prime}$ represent their post-collision velocities, $|\mathbf{g}|$ is the magnitude of their relative velocity, and $\mathbf{\omega}$ is the unit vector defining the post-collision relative velocity direction. The vector $\mathbf{\omega}$ performs a random walk on a unit sphere, which depends on collision models. For variable hard sphere(VHS) model, $\mathbf{\omega}$ is given as
	\begin{equation}
		\mathbf{\omega}={{\left( \cos \chi ,\sin \chi \cos \theta ,\sin \chi \sin \theta  \right)}^{T}},
	\end{equation}
	where, in polar coordinates, the cosine of deflection angle $\chi$ and azimuth angle $\theta$ are uniformly distributed over the intervals $[-1, 1]$ and $[0, 2\pi]$, respectively. That is,
	\begin{equation}
		\cos \chi =2{{\mathscr{R}}_{1}}-1,\text{ }\theta =2\pi {{\mathscr{R}}_{2}}.
	\end{equation}

    \subsection{Stochastic particle Shakhov method}~{}
	\label{SP Shakhov}
	For brevity, we focus exclusively on reviewing the BGK-type stochastic method in this section. The BGK-type model approximates the collision term in the Boltzmann equation (Eq.(\ref{Boltzmann equation})) using a simplified relaxation term, which is in the following form
	\begin{equation}
		\frac{\partial f}{\partial t}+\mathbf{c}\cdot \frac{\partial f}{\partial \mathbf{r}}={{J}_\mathrm{BGK}}=\frac{ g-f }{\tau}. \label{Boltzmann-BGK equation}
	\end{equation}
	In BGK-type collision model, a distribution function $f$ relaxes towards a target distribution function $g$ with a relaxation time $\tau$. Here, the relaxation time $\tau$ is defined as ${\mu} /{p}$, where $\mu$ and $p$ are the temperature-dependent dynamic viscosity and static pressure, respectively. The classical BGK model approximates the target distribution function $g$ by the Maxwellian distribution function $f^M$,
	\begin{equation}
		{{f}^{M}}=\frac{\rho }{{{\left( 2\pi RT \right)}^{3/2}}}\exp \left( -\frac{{{\mathbf{C}}^{2}}}{2RT} \right),
	\end{equation}
	where $\rho$, $\mathbf{U}$, ${T}$, $R$ and $\mathbf{C}$ denote the density,  macroscopic velocity, thermodynamic temperature, specific gas constant, and peculiar velocity defined as $\mathbf{c}-\mathbf{U}$, respectively. In contrast to classical BGK model, the Shakhov model explicitly modifies the target distribution function $g^S$ to correct the heat flux and achieve the proper Prandtl number. This modified distribution function $g^S$ takes the form
	\begin{equation}
		{{g}^{S}}={{f}^{M}}\left[ 1+(1-\Pr )\frac{\mathbf{Q}\cdot \mathbf{C}}{5pRT}\left( \frac{{{\mathbf{C}}^{2}}}{2RT}-5 \right) \right], \label{shakhov}
	\end{equation}
	where Pr is the Prandtl number, and $\mathbf{Q}$ denotes the heat flux vector, defined as
	\begin{equation}
		\mathbf{Q}=\int{\frac{1}{2}m\mathbf{C}{{\mathbf{C}}^{2}}fdc},
	\end{equation}
	where $m$ is the mass of a single gas molecule.

	In the remainder of this section, the stochastic particle method based on the Shakhov model is briefly reviewed. Similar to the DSMC method, the SP-Shakhov method also employs the operator splitting scheme, where Eq.(\ref{Boltzmann-BGK equation}) is decomposed into two sequential steps: free-transport step and relaxation step, i.e.,
	\begin{equation}
		\begin{aligned}
		& \frac{\partial f}{\partial t}+\mathbf{c}\cdot \frac{\partial f}{\partial \mathbf{r}}=0 \quad \text { and } \\
		& \frac{\partial f}{\partial t}=\frac{1}{\tau}\left(g^S-f\right).
		\end{aligned}
		\label{Shakohv-equation}
	\end{equation}
	The relaxation process due to particle collisions is mathematically described by the integral solution of the homogeneous Shakhov equation, and can be discretized for implementation at each time step $\Delta t$ as
	\begin{equation}
		f(t+\Delta t)={f}(t+\Delta t){{e}^{-\Delta t/\tau }}+\left( 1-{{e}^{-\Delta t/\tau }} \right)g^S(t). \label{integral solution}
	\end{equation}

	Similar to the DSMC method, the distribution function is approximated using a finite number of simulated particles. As derived from Eq.(\ref{integral solution}), a fraction ${e}^{-\Delta t/\tau }$ of particles undergo free molecular transport while retaining their velocities, with the remaining $(1-{e}^{-\Delta t/\tau })$ fraction undergoing velocity re-sampling from the Shakhov equilibrium distribution function $g^S$. According to the Eq.(\ref{Shakohv-equation}), the time evolution of particles is split into two consecutive steps: First, all particles within each computational cell are accurately tracked according to Eq.(\ref{free transport}), following the standard DSMC free-transport procedure. Then, each particle relaxes to the shakhov distribution function $g^S$ with a probability of $P^\mathrm{SBGK}$, which is given as
	\begin{equation}
		P^\mathrm{SBGK}=1-\exp \left( -\Delta t/\tau  \right). \label{Ncol_SBGK}
	\end{equation}

	The AR method is employed to sample post-relaxation particle velocities from the Shakhov distribution \cite{pfeiffer2018particle}. Because the Shakhov distribution function decomposes into a Maxwellian distribution multiplied by a polynomial function, velocity sampling proceeds in two stages: (1) Maxwellian sampling. Thermal velocity $\mathbf{C}$ components are drawn from the Maxwellian distribution using conventional direct sampling methods, i.e.,
	\begin{equation}
		\begin{aligned}
			C_x & =\sqrt{2 R T} \cos \left(2 \pi \mathscr{R}_1\right) \sqrt{-\ln \left(\mathscr{R}_2\right)}, \\
			C_y & =\sqrt{2 R T} \cos \left(2 \pi \mathscr{R}_3\right) \sqrt{-\ln \left(\mathscr{R}_4\right)}, \\
			C_z & =\sqrt{2 R T} \sin \left(2 \pi \mathscr{R}_3\right) \sqrt{-\ln \left(\mathscr{R}_4\right)}.
		\end{aligned} \label{Maxwell}
	\end{equation}
	(2) polynomial weighting and AR test. The polynomial function is extracted from Eq.(\ref{shakhov}) and expressed as the weighting factor $A$ (Eq.(\ref{A_cal})). Using the sampled thermal velocity $\mathbf{C}$, compute A. The envelope function $A_{max}$ is then evaluated by applying the same heat flux $\mathbf{Q}$ but with artificially large velocity magnitudes (e.g., $\left|\mathbf{C}\right|=5\sqrt{RT}$). Finally, the AR test is executed by calculating $A/A_{max}$. If $A/A_{max}>\mathscr{R}_5$, the thermal velocity $\mathbf{C}$ is accepted, and the post-relaxation particle velocity is $\mathbf{C}+\mathbf{U}$. Otherwise, the value of thermal velocity $\mathbf{C}$ is resampled until the condition is satisfied.
	\begin{equation}
		A = 1+(1-\Pr )\frac{\mathbf{Q}\cdot \mathbf{C}}{5pRT}\left( \frac{{{\mathbf{C}}^{2}}}{2RT}-5 \right).
		\label{A_cal}
	\end{equation}
	To enforce momentum and energy conservation, all particle velocities are rescaled following the relaxation procedure. The resultant particle velocities ${\mathbf{c}^{'}}$ are calculated as
	\begin{equation}
		{{\mathbf{c}}^{'}}=\mathbf{U}+\left( {{\mathbf{c}}_{p}}-{{\mathbf{U}}_{p}} \right)\sqrt{\frac{T}{{{T}_{p}}}},
	\end{equation}
	where $\mathbf{U}$ and $T$ donate average velocity and temperature before relaxation, ${\mathbf{U}}_{p}$ and $T_p$ represent post-relaxation average velocity and temperature, ${\mathbf{c}}_{p}$ is post-relaxation velocity of individual particles.

    \section{The particle Shakhov-DSMC hybrid method}\label{hybrid method}
	\subsection{The coupling strategy of hybrid method}~{}
	A multi-scale Boltzmann equation was recently derived in Ref.\cite{liu2024multi}, accompanied by the proposal of a novel wave-particle method. By employing the temporal integral solution of homogeneous BGK equation (Eq.(\ref{integral solution 2})), which serves as an alternative expression of the Second Law of Thermodynamics and remains independent on physical scales, this multi-scale Boltzmann equation can be derived as Eq.(\ref{MBE}).
    \begin{equation}
		f(t+\Delta t)=e^{-\frac{\Delta t}{\tau}} f(t)+\left(1-e^{-\frac{\Delta t}{\tau}}\right) g(t). \label{integral solution 2}
	\end{equation}
	\begin{equation}
		\frac{Df}{Dt}={{e}^{-\Delta t/\tau }}J_\mathrm{Boltzmann}+\left( 1-{{e}^{-\Delta t/\tau }} \right)L(f), \label{MBE}
	\end{equation}
	where $L(f)$ represents the linearized Boltzmann equation,valid under near-equilibrium conditions. Consequently, the multi-scale Boltzmann equation can be interpreted as a weighted average of non-equilibrium Boltzmann one and the linearized (near-equilibrium) Boltzmann one. These weights are determined by the observation time scale $\Delta t$, as defined in Eq.(\ref{integral solution 2}). In this framework, the introduction of the observation scale transforms the (master) Boltzmann equation into a multi-scale formulation.

	Since the BGK model equation represents a first-order approximation of linearized Boltzmann equation, the multi-scale Boltzmann equation can be approximated as follows
	\begin{equation}
		\frac{Df}{Dt}={{e}^{-\Delta t/\tau }}J_\mathrm{Boltzmann}+\left( 1-{{e}^{-\Delta t/\tau }} \right)J_\mathrm{BGK}. \label{MBE1}
	\end{equation}
	From the perspective of numerical methods, the observation time in this work corresponds to the local time step of a computational cell. Meanwhile, for stability considerations, a reasonable approximation of exponential weights can be adopted, leading to the following equation (see Eq.(\ref{MBE2})), where $P=1-{\tau}/{\Delta t}$. For practical implementation, $P$ can also be designed as a piecewise continuous function, defined by Eq.(\ref{P}).	
	\begin{equation}
		\frac{Df}{Dt}=\left( 1-P \right)J_\mathrm{Boltzmann}+PJ_\mathrm{BGK}. \label{MBE2}
	\end{equation}
	\begin{equation}
		P=\left\{\begin{array}{ll}0,&\Delta t\leq\tau,\\1-\frac{\tau}{\Delta t},&\Delta t>\tau.\end{array}\right.
	\label{P}
	\end{equation}
    
	For a particle-based method, this multi-scale collision term (see Eq.(\ref{MBE2}))can be interpreted as a two-step process. First, the Boltzmann collision term $J_\mathrm{Boltzmann}$ is solved using the DSMC method. Second, the BGK-type collision term is handled via the stochastic-particle Shakhov method. The evolution time scaling ratio $P$ between the two collision terms is determined by Eq.(\ref{P}). Therefore, according to the Eq.(\ref{MBE2}), the rarefied DSMC collision term and near-continuum Shakhov collision term are automatically coupled in the present hybrid Shakhov-DSMC method, which enlarges the time step of DSMC while improving the accuracy of the SP Shakhov method.

	In the hybrid Shakhov-DSMC method, significant mesh variations often arise during numerical implementation. This necessitates establishing a quantitative relationship between the observation time step $\Delta t$ and the global (advancing) time step $\Delta t_G$. Here, the observation time step $\Delta t$ is determined by the local time step $\Delta t_c$ of each computational cell, where
	\begin{equation}
		\Delta t_{c,k}=\frac{L_\mathrm{cell}}{\left| \mathbf{U}_k \right|+\sqrt{\gamma RT_k}}, \label{local time step}
	\end{equation}
	where the subscript ``$k$'' denotes the cell index, $\mathbf{U}_k$ is the macroscopic velocity of cell $k$, $\gamma$ is the specific heat ratio, and $L_\mathrm{cell}$ can be selected flexibly to match the grid resolution, which is set to the minimum cell length $L_{min}$ in this work. Crucially, the simulation advances globally using a time step $\Delta t_G$, determined conservatively as the minimum among all local characteristic time steps $\Delta t_{c,k}$ across the computational domain, i.e,
	\begin{equation}
		\Delta t_G=\min \left( \Delta t_{c,k} \right). \label{global time step}
	\end{equation}
	Note that hybrid methods\cite{liu2020unified, liu2020simplified, liu2024multi} allow the selection of multi-scale time step exclusively based on the Courant-Friedrichs-Lewy (CFL) condition, without being constrained by molecular mean collision time. However, the stochastic particle method introduces additional numerical viscosity in the continuum/near-continuum regimes due to its upwind nature of free transport\cite{fei2021hybrid}. Consequently, this results in degraded solution accuracy and imposes stricter constraints on the allowable time step for precise computations. (For potential solutions to this issue, please refer to relevant work in unified stochastic particle\cite{fei2020unified, chen2020three, fei2022unified}, which is beyond the scope of the present study.)
	
\begin{figure}
		\centering
		\subfigure[$\Delta {{t}_{c}}<\tau$]{\label{Classification-2}\includegraphics[width=0.35\textwidth]{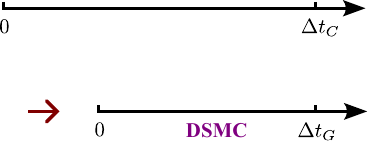}}
		\qquad
		\qquad
		\subfigure[$\Delta {{t}_{c}}>\tau$]{\label{Classification-1}\includegraphics[width=0.35\textwidth]{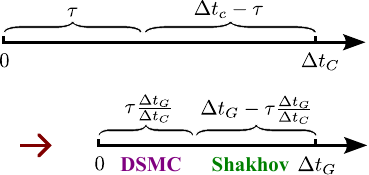}}
		\caption{\label{weight}Classification of global timestep and weight in hybrid method.}
	\end{figure}

	It should be designed the numerical hybrid method to match $\Delta t_c$ and $\Delta t_G$ in the process of coupling. More precisely, $\Delta t_c$ determines the physical scale, while $\Delta t_G$ evolves the flow field. In general, there are two situations, as shown in Fig.\ref{weight}. (1) When $\Delta t_c\leq\tau$, the hybrid solver becomes a pure DSMC one. (2) When $\Delta t_c>\tau$, both hybrid Shakhov and DSMC solver are used. In the time period $\Delta t_c$, single collision happens in $\left[0, \tau \right)$, multiple collision happens in $\left[\tau, \Delta t_c\right)$. However, the numerical simulation goes forward by global time step $\Delta t_G$, so a narrowing is used ($\Delta t_G / \Delta t_c < 1 $). The two time periods forward are narrowed proportionally into $\left[0, \tau \frac{\Delta t_G}{\Delta t_c}\right)$, and $\left[\tau \frac{\Delta t_G}{\Delta t_c}, \Delta t_G\right)$, as illustrated in Fig.\ref{weight}.

	In summary, the collision time $\Delta t_{c,\mathrm{DSMC}}$ for DSMC in hybrid method is
	\begin{equation}
		\Delta t_{c,\mathrm{DSMC}}=\left\{\begin{array}{ll}\Delta t_G,&\Delta t_c\leq\tau,\\\frac{\tau}{\Delta t_c}{\Delta t_G},&\Delta t_c>\tau.\end{array}\right. \label{tDSMC}
	\end{equation}
	and the relaxation time $\Delta t_{c,\mathrm{SBGK}}$ of stochastic particle Shakhov method is
	\begin{equation}
		\Delta t_{c,\mathrm{SBGK}}=\left\{\begin{array}{ll} 0,&\Delta t_c\leq\tau,\\ \left(1-\frac{\tau}{\Delta t_c}\right){\Delta t_G},&\Delta t_c>\tau.\end{array}\right. \label{tShakhov}
	\end{equation}
	The collision time $\Delta t_{c,\mathrm{DSMC}}$ and $\Delta t_{c,\mathrm{SBGK}}$ are used to calculate the number of particle pairs in Eq.(\ref{Ncol_DSMC}) and relaxation in Eq.(\ref{Ncol_SBGK}), respectively. The detailed algorithms for these two collision terms are almost the same with the ones for the pure DSMC and pure SP methods, which can be found in Section \ref{DSMC} and Section \ref{SP Shakhov}.

    \subsection{The implementation of hybrid strategy}
\begin{figure}
		\centering
		\includegraphics[width=0.7\textwidth]{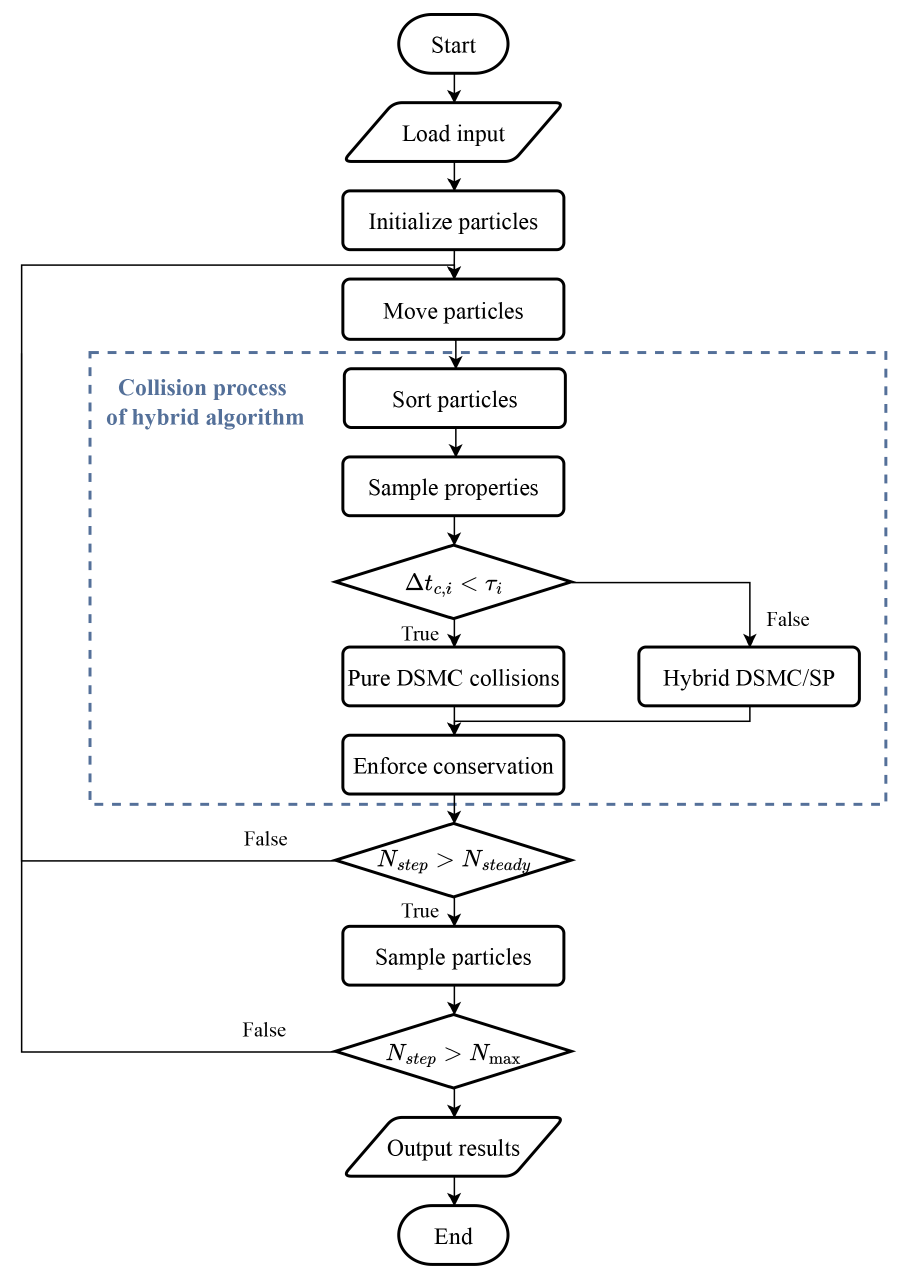}
		\caption{\label{flowchart} Flow chart of hybrid Shakhov-DSMC solution procedure.}
	\end{figure}

~{}
	Using the operator splitting scheme, the Eq.(\ref{MBE}) is divided into free-transport and collision steps, i.e.,
	\begin{equation}
		\begin{aligned}
		& \frac{\partial f}{\partial t}+\mathbf{c}\cdot \frac{\partial f}{\partial \mathbf{r}}=0 \quad \text { and } \\
		& \frac{\partial f}{\partial t}=\left( 1-P \right)J_\mathrm{Boltzmann}+PJ_\mathrm{BGK},
		\end{aligned}
	\end{equation}
	where the Boltzmann collision integral $J_\mathrm{Boltzmann}$ is computing using the DSMC method, while the BGK-type collision term $J_\mathrm{BGK}$ is handled via SP shakhov method. Similar to DSMC and SP Shakhov methods, the hybrid algorithm consists of four primary phases: initialization, particle movement, collision processing and statistical sampling. Fig.\ref{flowchart} presents a comprehensive flowchart of the hybrid Shakhov-DSMC algorithm. The complete computational procedure can be summarized as follows:

	\textbf{Step(1):} Initialization. The simulation input data is read and the simulated particles are initialized within the computational domain with positions ${\mathbf{r}}^{0}_{i}$ and velocities ${\mathbf{c}}^{0}_{i}$. Positions ${\mathbf{r}}^{0}_{i}$ are randomly distributed throughout the computational domain. Velocities ${\mathbf{c}}^{0}_{i}$ are initialized as ${\mathbf{C}}^{0}_{i}+{\mathbf{U}}^{0}_{i}$, where ${\mathbf{C}}^{0}_{i}$ donates the thermal velocity component sampled from a Maxwellian distribution (see Eq.(\ref{Maxwell})), and ${\mathbf{U}}^{0}_{i}$ is the initial macroscopic velocity of the flow field.

	\textbf{Step(2):} Particle movement. According to Eq.\ref{free transport}, all simulated particles are moved to their new positions, with boundary conditions (diffuse or specular reflection) applied when particle-wall interactions occur. Particles leaving the computational domain are permanently removed from the simulation. For diffuse reflections with full thermal accommodation employed in this work, the normal velocity $c_n$ is sampled from
	\begin{equation}
		c_n=\sqrt{2RT_{wall}}\cos \sqrt{-\ln \left( \mathscr{R}_1 \right)}.
	\end{equation}
	The tangential components $c_s$ and $c_t$ are sampled according to those $y$ and $z$ at the wall temperature $T_{wall}$ in Eq.(\ref{Maxwell}).

	\textbf{Step(3):} Particle indexing and macroscopic variables sampling. All simulated particles are spatially indexed into computational grid cells to enable efficient neighbor identification for collision pair selection. Meanwhile, the macroscopic flow variables within each cell—including number density $n_k$, macroscopic velocity $\mathbf{U}_{k}$, temperature $T_k$, and heat flux $\mathbf{Q}_{k}$—are calculated through ensemble averaging of particle properties. Theses sampled quantities serve as the governing parameters for velocity redistribution during Shakhov relaxation. The temperature $T_k$ and heat flux $\mathbf{Q}_{k}$ are particularly treated using unbiased estimators calculated as follows
	\begin{equation}
		\begin{aligned}
			& T=\frac{1}{N-1}\sum\limits_{i=1}^{N}{\frac{m\mathbf{C}_{i}^{2}}{3{{k}_{B}}}}, \\
			& \mathbf{Q}=\frac{1}{V}\frac{N}{(N-1)(N-2)}\sum\limits_{i=1}^{N}{\frac{1}{2}m}{\mathbf{C}}_{i}^{2}{{\mathbf{C}}_{i}}.
		\end{aligned}
	\end{equation}

	\textbf{Step(4):} Particle collision. The collision procedure follows a two-step process within each computational cell. The local observation time $\Delta t_c$ and relaxation time $\tau$ are computed for every cell, and the global time step is determined $\Delta t_G$ according to Eq.(\ref{local time step}) and Eq.(\ref{global time step}). When $\Delta t_c\leq\tau$, the pure DSMC solver is used, with the collision time $\Delta t_{c,\mathrm{DSMC}}$ set equal to $\Delta t_G$. Particle velocities are updated using the DSMC algorithm (section \ref{DSMC}). When $\Delta t_c>\tau$, the solver switches to a hybrid approach, where $\Delta t_{c,\mathrm{DSMC}}$ (collision time) and $\Delta t_{c,\mathrm{SBGK}}$ (relaxation) time are calculated from Eq.(\ref{tDSMC}) and Eq.(\ref{tShakhov}), respectively. The velocities of the simulated particles are updated according to the corresponding collision algorithm in section \ref{DSMC} and section \ref{SP Shakhov}. To ensure momentum and energy conservation, all particle velocities—whether colliding, relaxing, or unchanged—are rescaled after the collision step. Step(2)-Step(4) of these procedures are repeated until the flow field reaches steady.

	\textbf{Step(5):} Sample the information of particles. Particle data are sampled once the flow reaches steady-state conditions ($N_{step}>N_{steady}$). The number density and post-collision velocities of simulated particles are accumulated to compute macroscopic flow properties. Within each computational cell, three statistical variables are recorded: the total particle count $\sum{N}$, the sum of particle velocities $\sum{\mathbf{c}}$, and the sum of squared velocities $\sum{{\mathbf{c}}^2}$. To  minimize statistical noise, Step(2)-Step(5) are repeated until the sampling size is large enough.

	\textbf{Step(6):} Output the results of flow field. The flow field results are output once the simulation reaches the final time step ($N_{step}>N_{max}$). The macroscopic flow properties are calculated by averaging the accumulated values of simulated particles from Step(5).

    \section{Numerical simulations}\label{Number simulations}
	In this section, five benchmark cases including 1-D sod shock tube, 2-D hypersonic flow around cylinder and jet expansion into the vacuum, 3-D hypersonic flows around sphere and X-38 like vehicle are simulated to validate the performance of the hybrid method. In all test cases, the flow is assumed to be monatomic argon gas. The variable hard sphere (VHS) model is employed with a reference diameter $d_{ref}=4.17\times{{10}^{-10}m}$ at $T_{ref}= 273 K$, and the reference viscosity is ${{\mu }_{ref}}=2.117\times {{10}^{-5}}Pa\cdot s$ with a power-law exponent $\omega =0.81$. All DSMC results are obtained by either DS2V\cite{2013The} or SPARTA\cite{plimpton2019direct} with no-time-counter (NTC) scheme, and hybrid method results are obtained by JiutianSolver. In all cases, the global Knudsen number $\mathrm{Kn}$ is defined as
	\begin{equation}
		\mathrm{Kn}=\frac{\lambda}{L_{ref}},
	\end{equation}
	where $L_{ref}$ is the reference length, $\lambda$ is the mean free path which is determined for VHS molecules, i.e.,
	\begin{equation}
		\lambda =\frac{1}{\sqrt{2}\pi d_{ref}^{2}n{{\left( \frac{{{T}_{ref}}}{T} \right)}^{\omega -0.5}}}.
	\end{equation}

    \subsection{Sod shock tube}~{}
	The sod shock tube is a typical non-equilibrium flow problem, which consists of two regions with distinct initial conditions separated by a diaphragm. When the diaphragm is removed, a shock wave and rarefaction wave are generated, and the flow evolves over time. In the case, upstream $\mathrm{Kn}$ numbers are taken from (${10}^{-2}$, ${10}^{-3}$ and ${10}^{-4}$), and the initial conditions are specified in Table \ref{Sod shock stube}. The computational domain spans $[{-0.5}m, {0.5}m]$, where the total domain length defines the reference length. At $\mathrm{Kn}=10^{-2}$, the cell size is $\lambda$ and computational domain is discretized into 100 uniform cells. At $\mathrm{Kn}={10}^{-3}$ and ${10}^{-4}$, the cell size is $4\lambda$ and computational domain is discretized into 250 and 2500 uniform cells, respectively. To minimize statistical fluctuations, each computational cell is initially populated with $10^5$ simulated particles. The left and right boundaries of computational domain are set as ghost cells which maintaining their initial values at each time step. The diaphragm is removed at $t=0$.

    \begin{table}
		\centering
		\caption{\label{Sod shock stube} The upstream and downstream parameters of Sod shock tube.}
		\begin{threeparttable}
			\begin{tabular}{p{60pt}<{\centering}|p{80pt}<{\centering}|p{50pt}<{\centering}|p{50pt}<{\centering}|p{80pt}<{\centering}|p{65pt}<{\centering}|p{50pt}<{\centering}}
				\hline
				\hline
				Case      &$\rho_{up}\ (kg/m^{-3})$ &$u_{up}\ (m/s)$ &$T_{up}\ (K)$ &$\rho_{down}\ (kg/m^{-3})$ &$u_{down}\ (m/s)$    &$T_{down}\ (K)$   \\
				\hline
				Kn=$10^{-2}$   & $8.58174 \times 10^{-6}$       & 0    & 273 
				& $1.07272 \times 10^{-6}$   & 0    & 218.4   \\
				\hline
				Kn=$10^{-3}$   & $8.58174 \times 10^{-5}$       & 0    & 273 
				& $1.07272 \times 10^{-5}$   & 0    & 218.4   \\
				\hline
				Kn=$10^{-4}$   & $8.58174 \times 10^{-4}$       & 0    & 273 
				& $1.07272 \times 10^{-4}$   & 0    & 218.4   \\
				\hline
				\hline
			\end{tabular}
		\end{threeparttable}		
	\end{table}

\begin{figure}
		\centering
		\subfigure[]{\label{Sod-Kn1E-2}\includegraphics[width=0.45\textwidth]{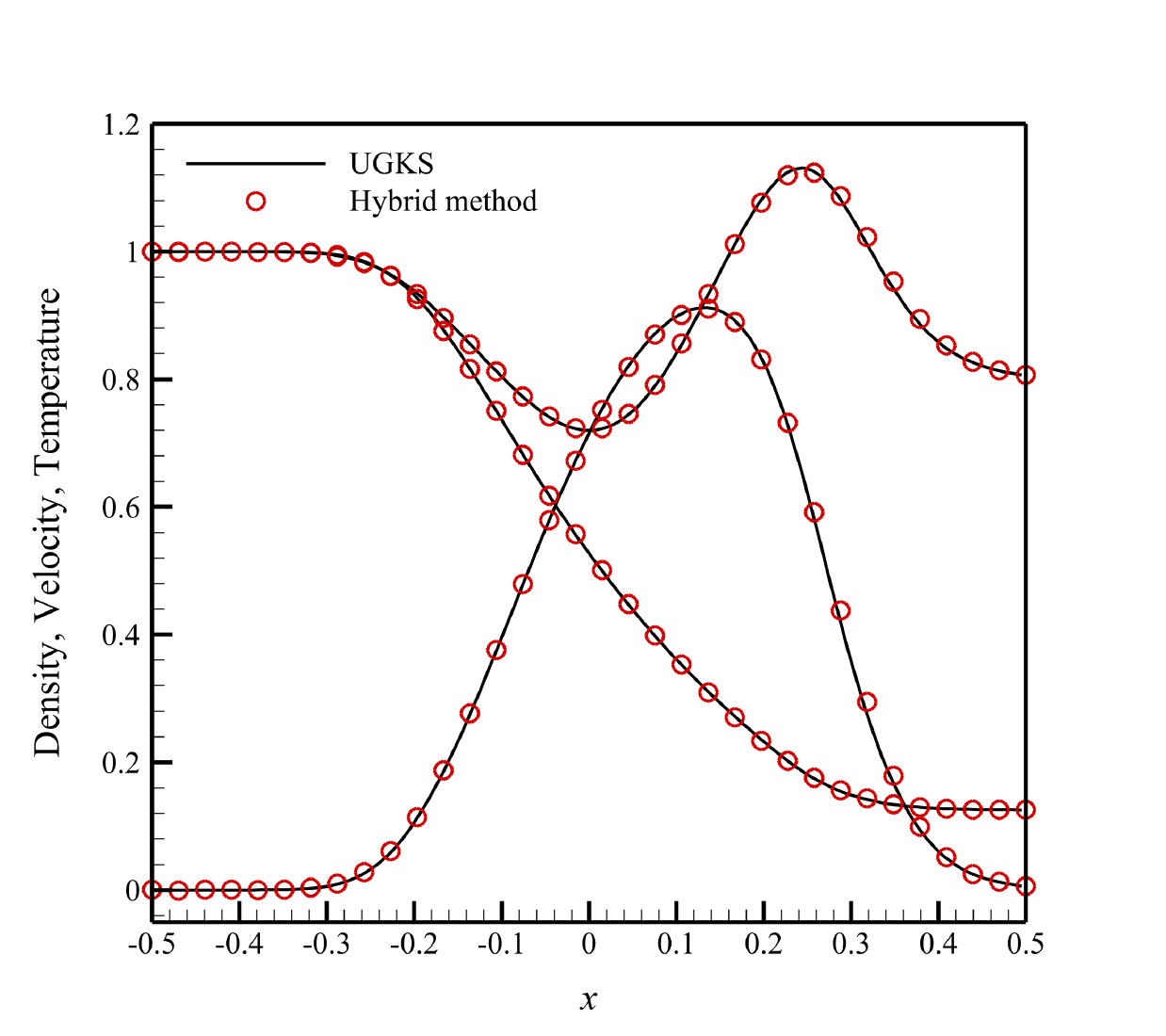}}
		\subfigure[]{\label{Sod-Kn1E-3}\includegraphics[width=0.45\textwidth]{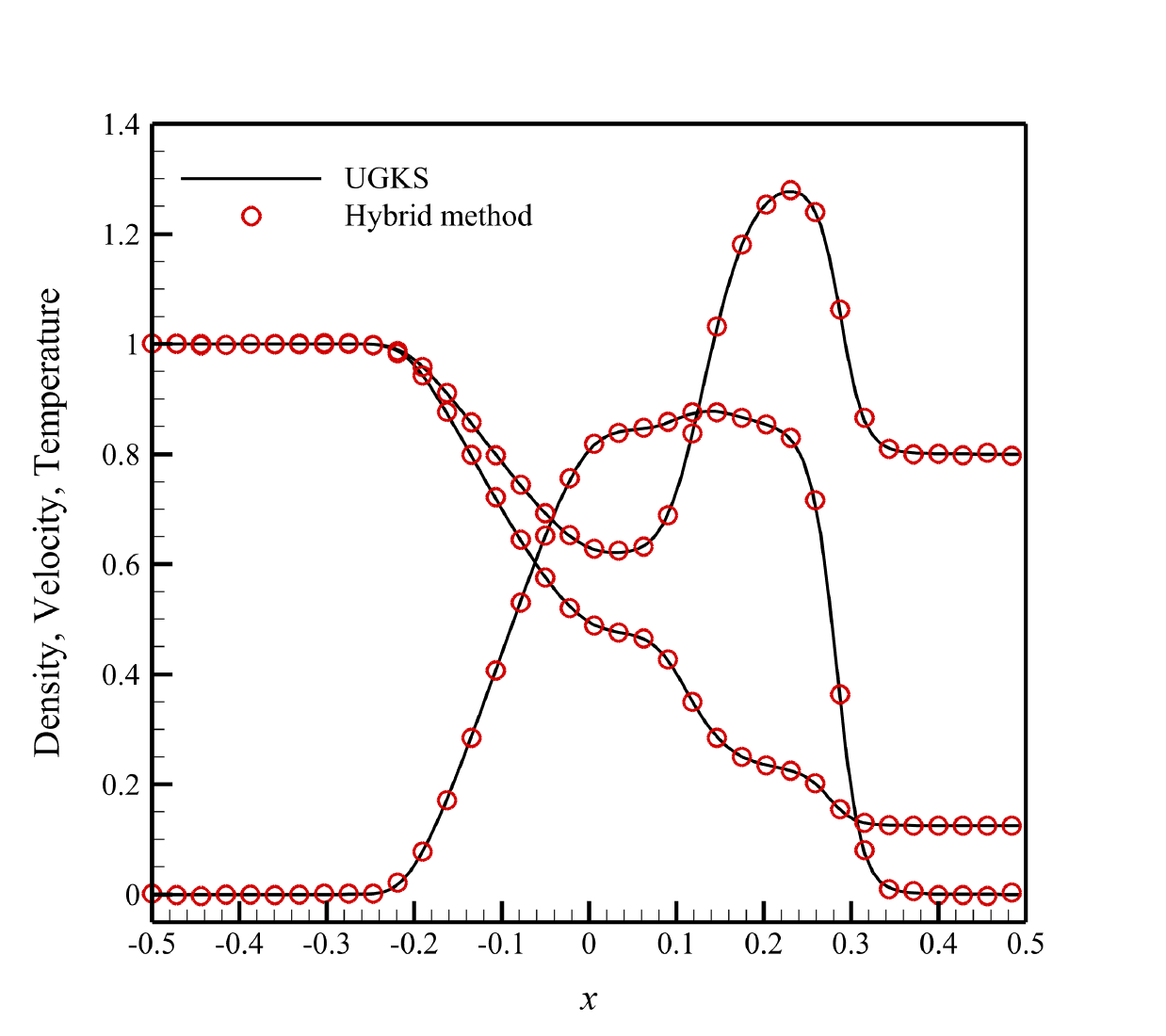}}
		\subfigure[]{\label{Sod-Kn1E-4}\includegraphics[width=0.45\textwidth]{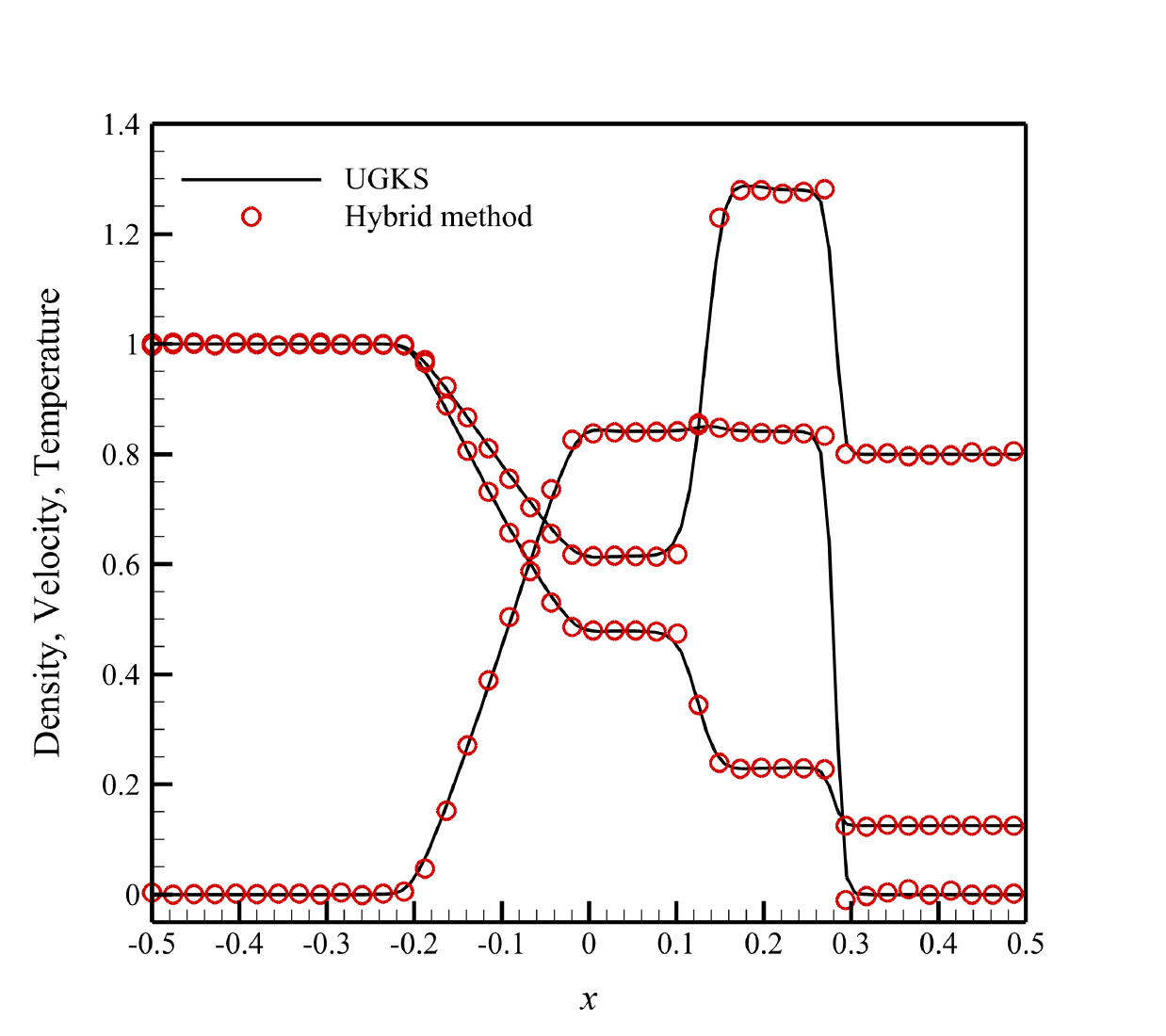}}
		\subfigure[]{\label{Sod-weight}\includegraphics[width=0.45\textwidth]{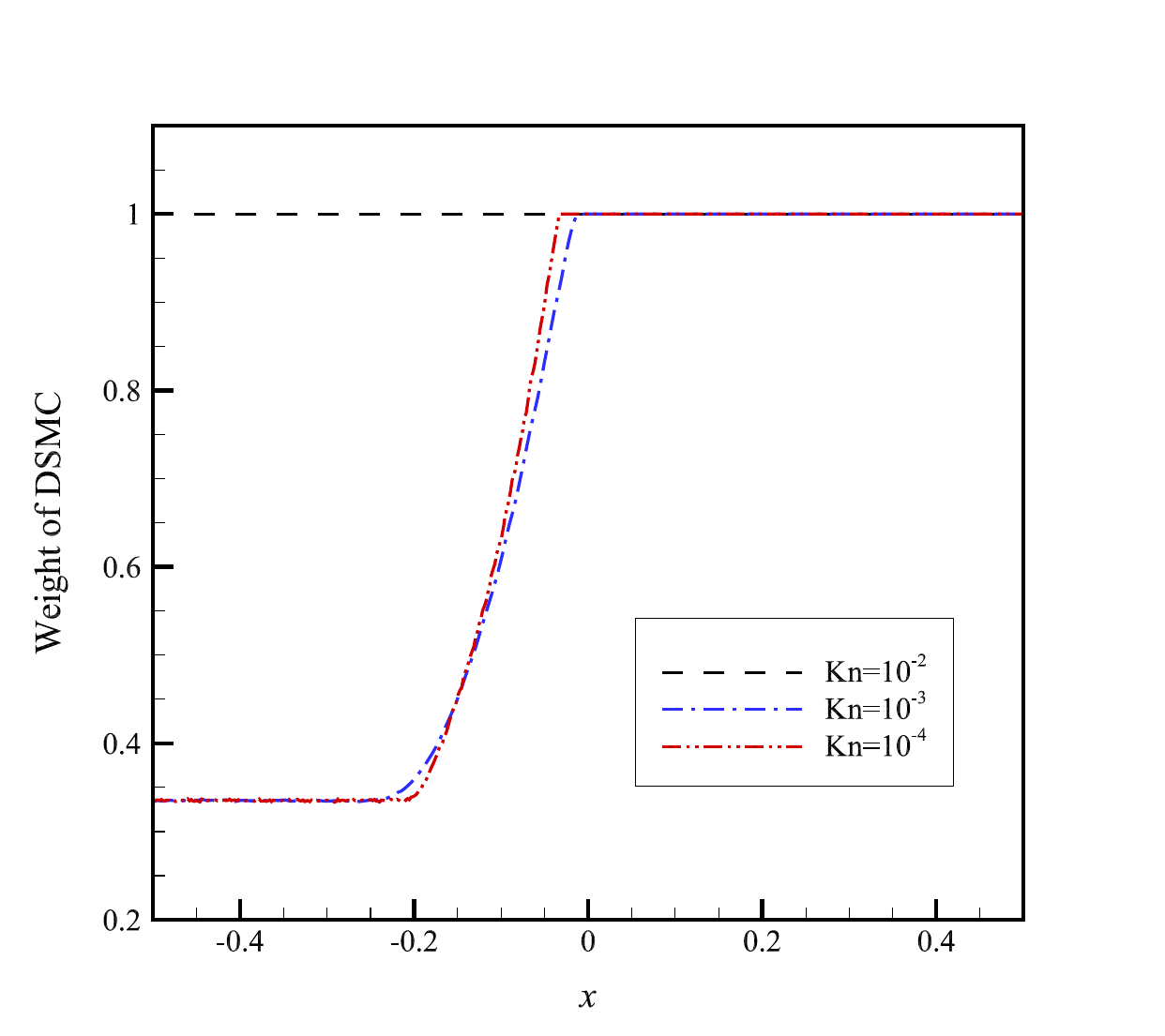}}
		\caption{\label{Sod}Profiles of sod shock tube at $t=0.15$: (a) $\mathrm{Kn}=10^{-2}$, (b) $\mathrm{Kn}=10^{-3}$, (c) $\mathrm{Kn}=10^{-4}$, (d) ratio of DSMC. Circle symbol: hybrid DSMC-Shakhov method; Solid line: UGKS results.}
	\end{figure}

    The dimensionless density, velocity and temperature profiles at $\tilde{t}=0.15$ for different Kn numbers are illustrated in Fig.\ref{Sod}, where all variables are normalized as follows
	\begin{equation}
		\tilde{\rho }=\frac{\rho }{{{\rho }_{up}}},\tilde{T}=\frac{T}{{{T}_{up}}},\tilde{u}=\frac{u}{\sqrt{R{{T}_{up}}}},\tilde{t}=\frac{t}{{{L}_{ref}}/\sqrt{R{{T}_{up}}}}. \label{dimensionless variables}
	\end{equation}
	The reference solutions are obtained using the unified gas-kinetic scheme (UGKS)\cite{xu2010unified}. Excellent agreement is observed between the hybrid method results and the reference solutions. Figure \ref{Sod-weight} presents the DSMC ratio variation across different $\mathrm{Kn}$ numbers. While the solution at $\mathrm{Kn}=10^{-2}$ employs a pure DSMC solver, the hybrid method increasingly utilizes the SP-Shakhov method as the flow becomes more continuum-dominated. However, at at $\mathrm{Kn}=10^{-3}$ and $\mathrm{Kn}=10^{-4}$, the DSMC ratios remain nearly identical, which can be explained by the uniform computational grid resolution ($\Delta x = 4 \lambda$).

    \subsection{Hypersonic flow past a cylinder}~{}
	The hypersonic flow past a cylinder is a typical case to investigate multi-scale gas flows. Numerical simulations are performed at a freestream Mach number of 5 for two Knudsen numbers, $\mathrm{Kn}=0.01$ and 0.1. The cylinder radius, chosen as the characteristic length, is set to $1m$. The freestream number densities are $1.29438 \times 10^{20} m^{-3}$ and $1.29438 \times 10^{19} m^{-3}$, corresponding to the two Knudsen numbers, while the freestream temperature and velocity are fixed at 273K and $1537.44 m/s$, respectively. A fully diffusive wall boundary condition is imposed, maintaining a surface temperature of 273K. The computational grid consists of 180 uniformly distributed nodes along the circumferential direction. The first mesh layer adjacent to the wall has a thickness of $0.001m$. For the case with $\mathrm{Kn}=0.01$, two meshes with different expansion rates are tested: mesh1 (growth rate = 1.1) and mesh2 (growth rate = 1.05). For the case with $\mathrm{Kn}=0.1$, a single mesh with a growth rate of 1.1 is employed.

\begin{figure}
		\centering
		\subfigure[]{\label{Ma5Kn001-weight1}\includegraphics[width=0.45\textwidth]{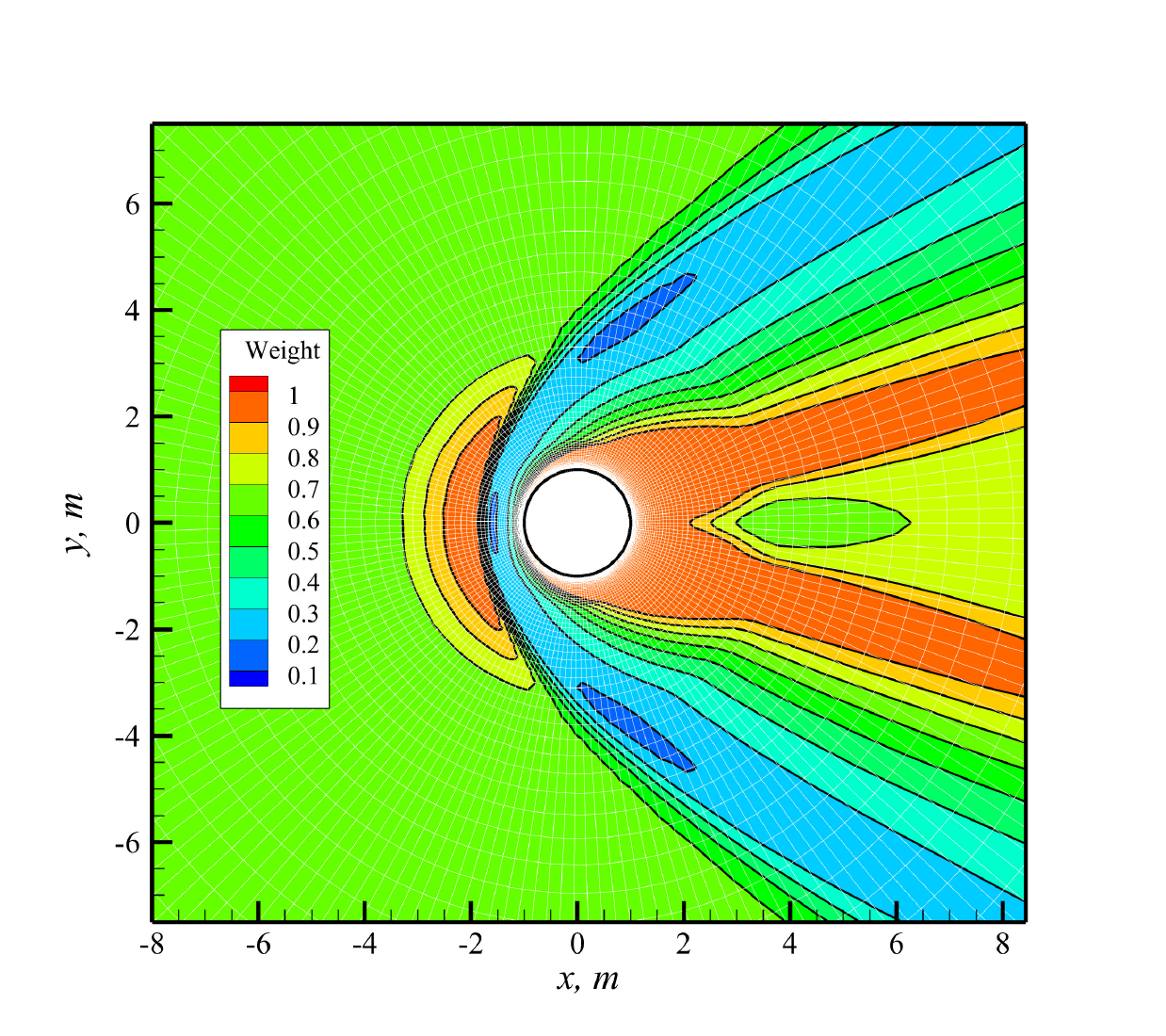}}
		\subfigure[]{\label{Ma5Kn001-weight2}\includegraphics[width=0.45\textwidth]{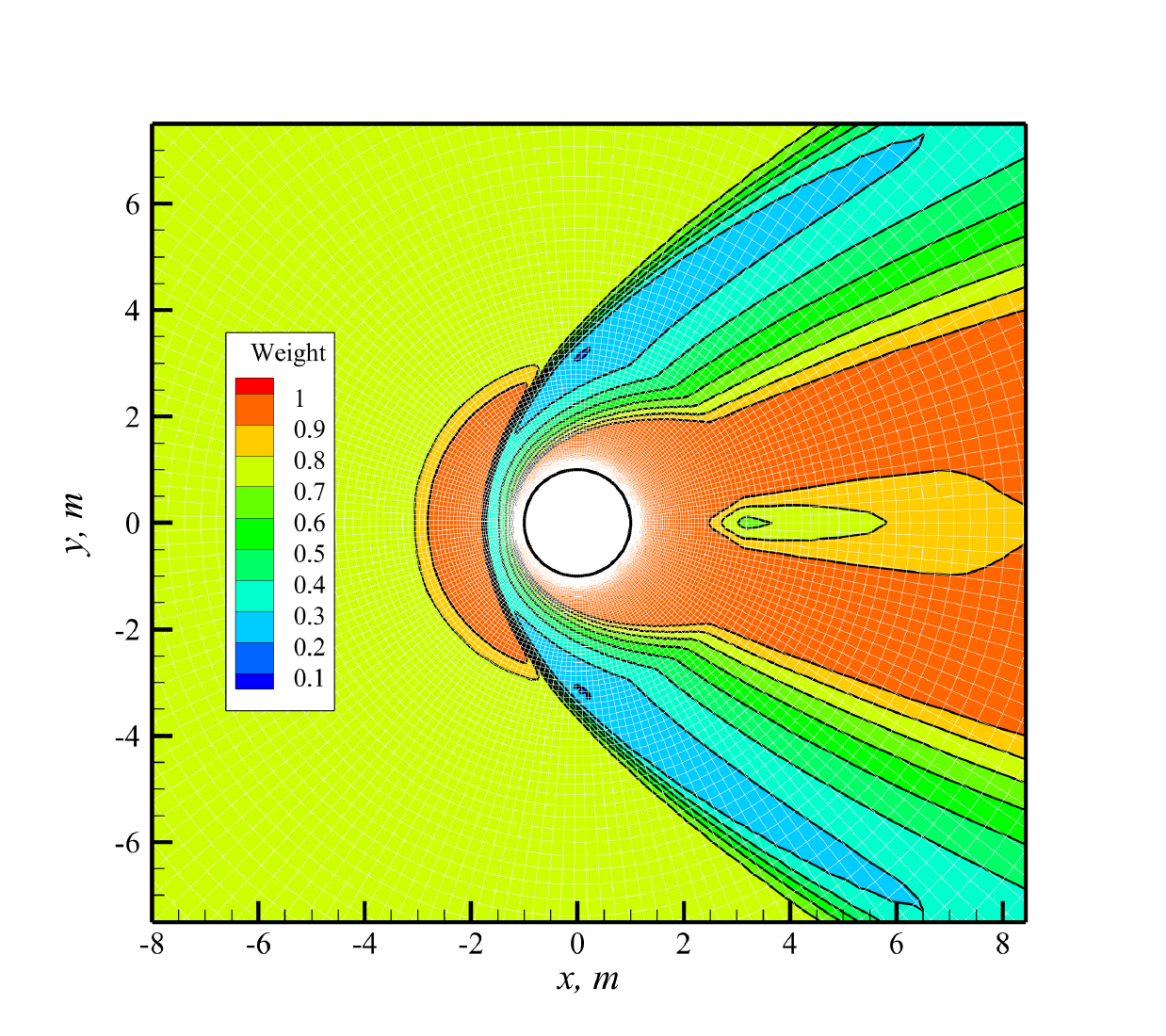}}
		\subfigure[]{\label{Ma5Kn001-zdx-temp}\includegraphics[width=0.45\textwidth]{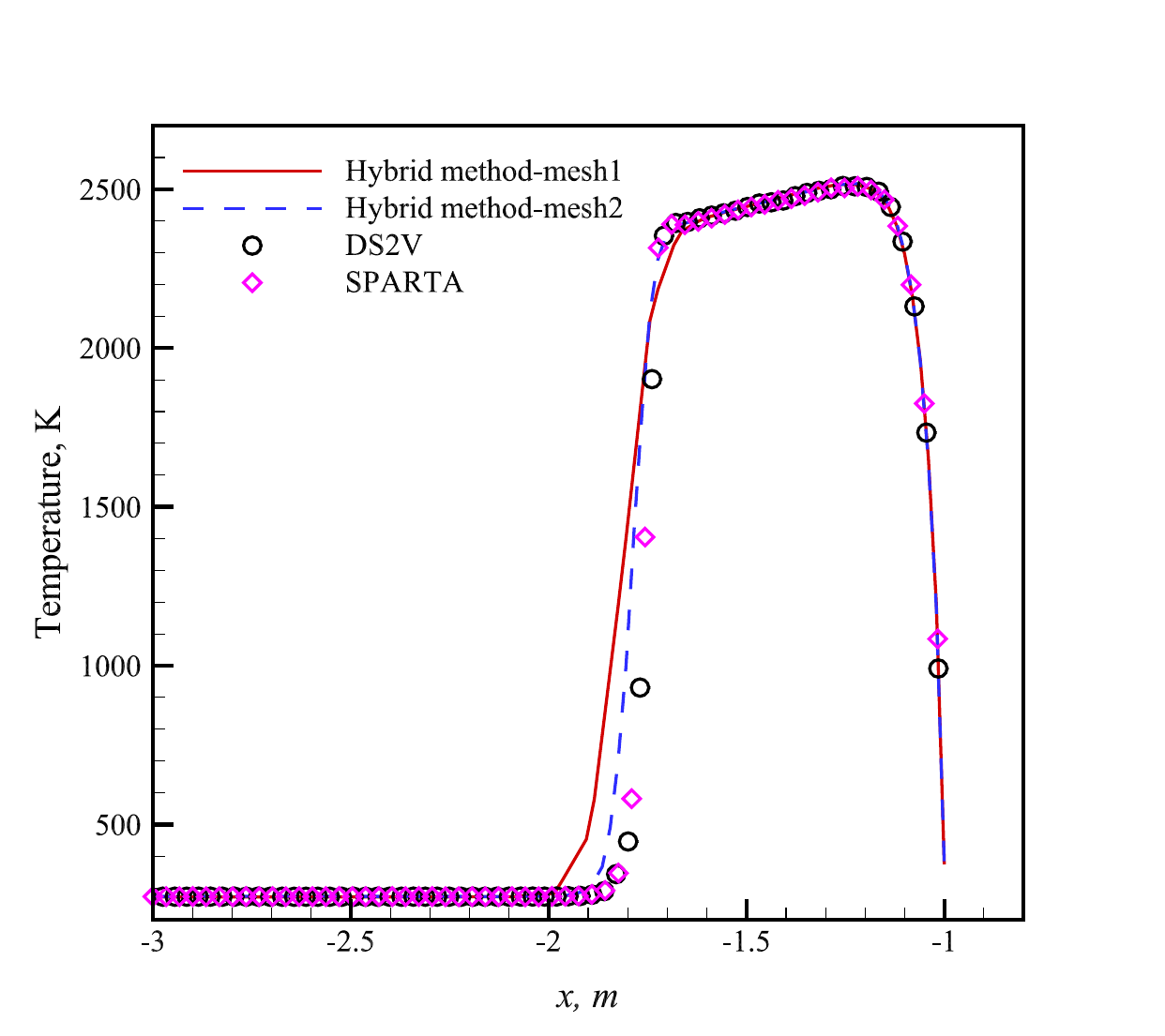}}
		\caption{\label{Cylinder-Ma5Kn001-flow} Results of hypersonic cylinder flow for $\mathrm{Ma}=5$, $\mathrm{Kn}=0.01$: (a) ratio of DSMC contour with computational mesh (white gridlines, mesh 1), (b) ratio of DSMC contour with computational mesh (white gridlines, mesh 2), (c) temperature along the stagnation line. Solid line: hybrid DSMC-Shakhov with mesh1; Dashed line: hybrid DSMC-Shakhov with mesh2, Circle symbol: DSMC from DS2V, Diamonds symbol: DSMC from SPARTA.}
	\end{figure}

\begin{figure}
		\centering
		\subfigure[]{\label{Ma5Kn001-pres}\includegraphics[width=0.45\textwidth]{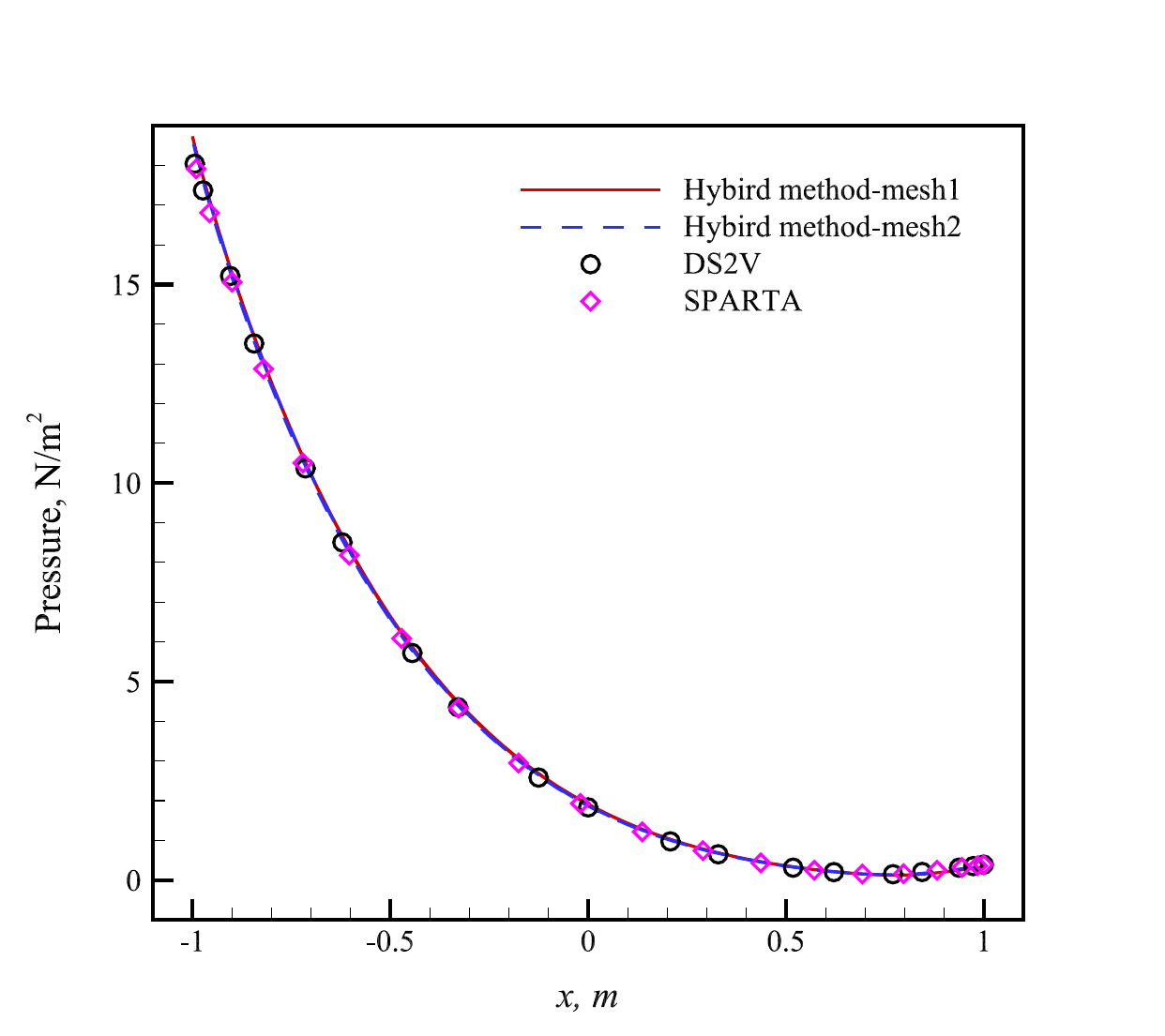}}
		\subfigure[]{\label{Ma5Kn001-tau}\includegraphics[width=0.45\textwidth]{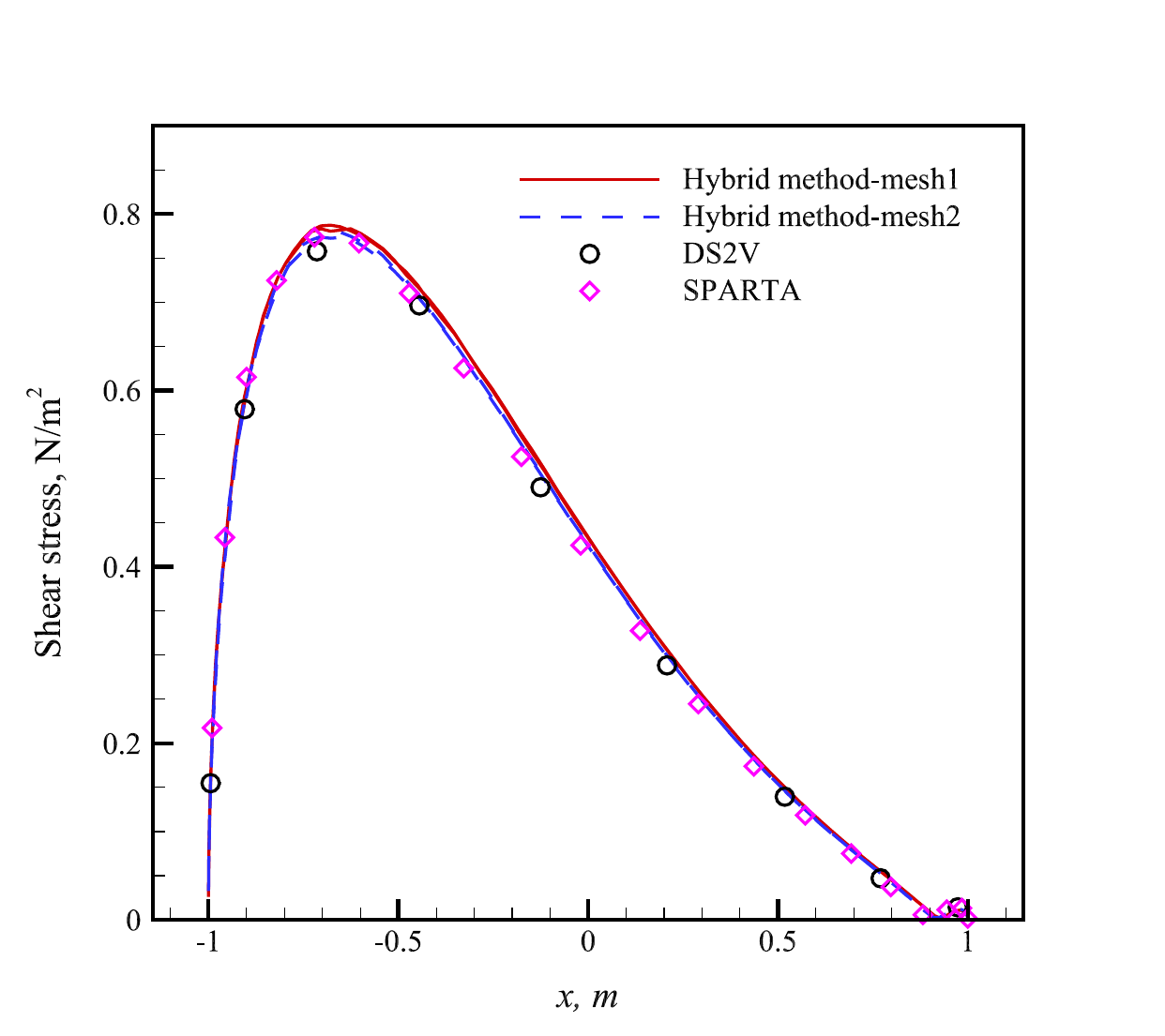}}
		\subfigure[]{\label{Ma5Kn001-q}\includegraphics[width=0.45\textwidth]{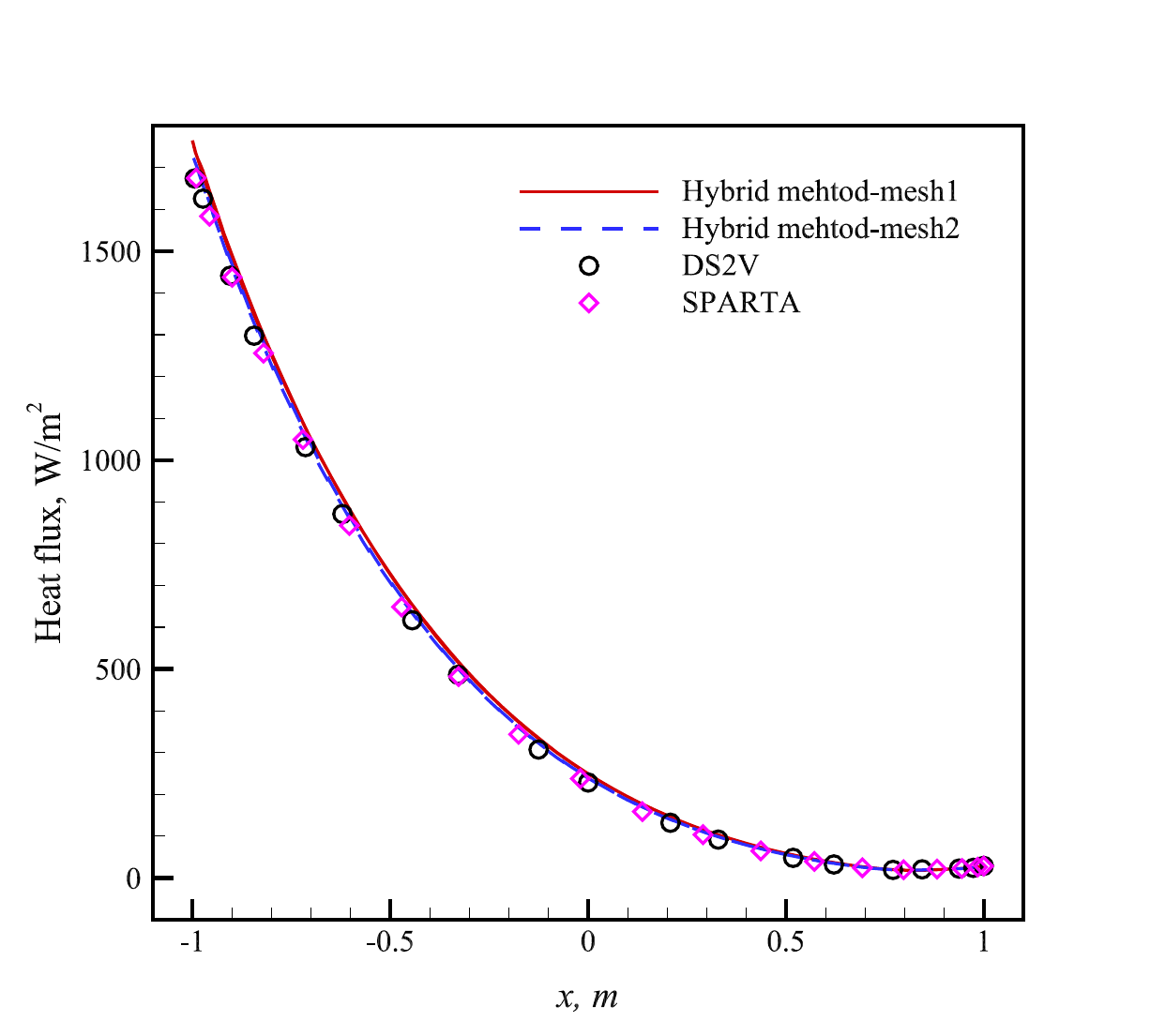}}
		\caption{\label{Cylinder-Ma5Kn001-surface} Surface results of hypersonic cylinder flow for $\mathrm{Ma}=5$, $\mathrm{Kn}=0.01$: (a) pressure, (b) shear stress, (c) heat flux. Solid line: hybrid DSMC-Shakhov with mesh1; Dashed line: hybrid DSMC-Shakhov with mesh2, Circle symbol: DSMC from DS2V, Diamond symbol: DSMC from SPARTA.}
	\end{figure}

\begin{figure}
		\centering
		\subfigure[]{\label{Ma5Kn01-temp}\includegraphics[width=0.45\textwidth]{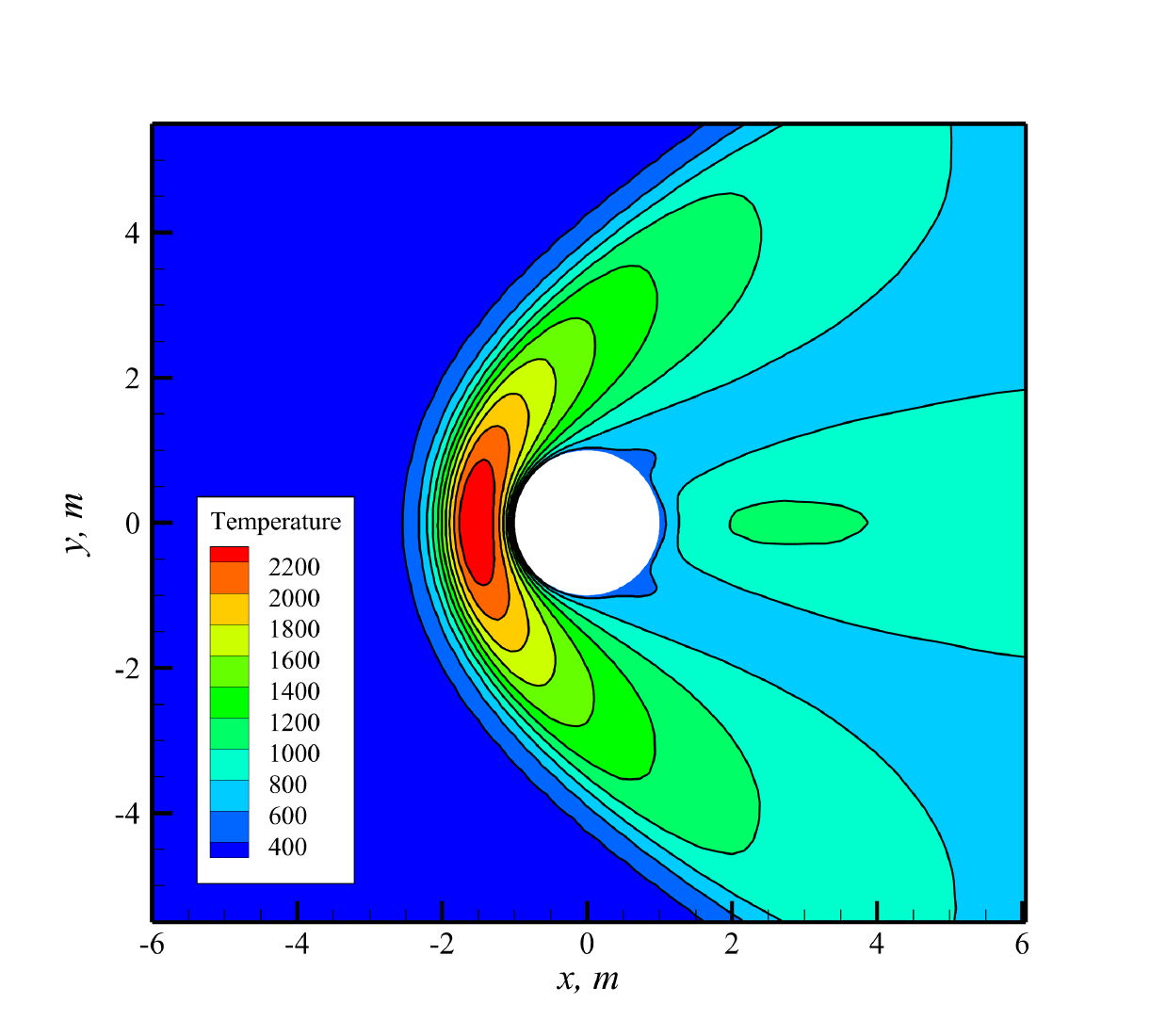}}
		\subfigure[]{\label{Ma5Kn01-weight}\includegraphics[width=0.45\textwidth]{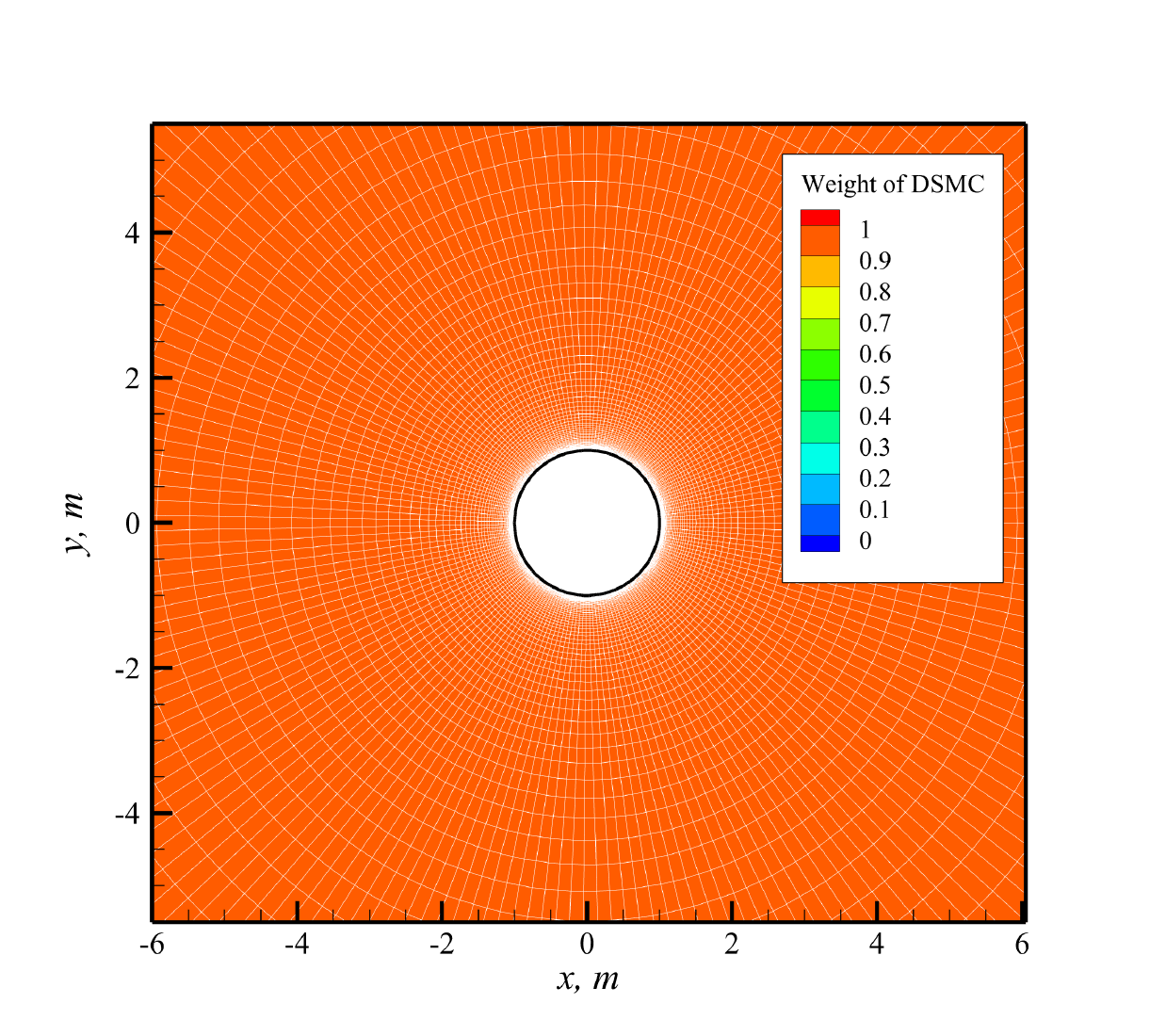}}
		\subfigure[]{\label{Ma5Kn01-zdx}\includegraphics[width=0.45\textwidth]{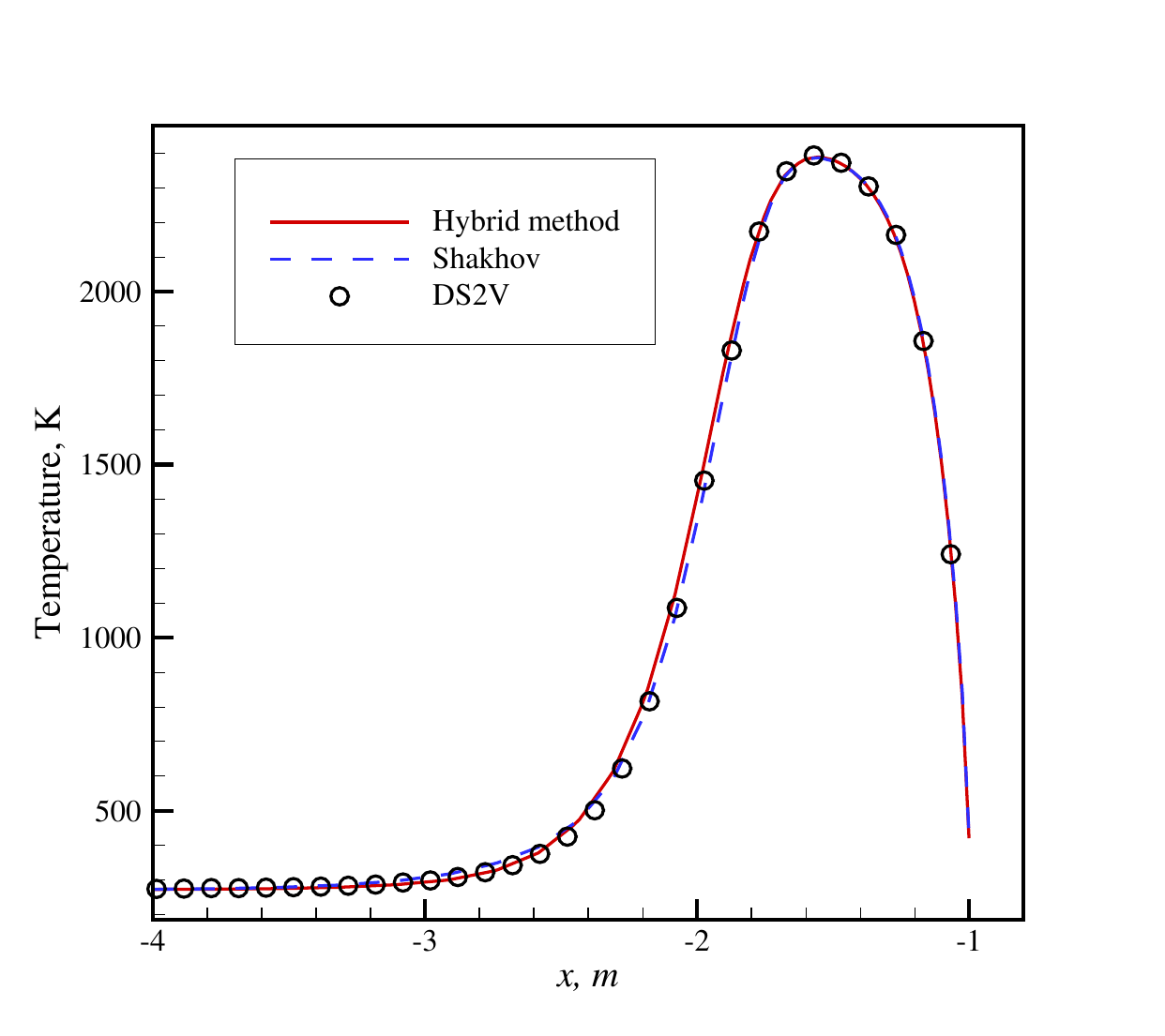}}
		\caption{\label{Cylinder-Ma5Kn01-flow} Results of hypersonic cylinder flow for $\mathrm{Ma}=5$, $\mathrm{Kn}=0.1$: (a) temperature contour, (b) ratio of DSMC contour with computational mesh (white gridlines), (c) temperature along the stagnation line.}
	\end{figure}

	Fig.\ref{Cylinder-Ma5Kn01-flow} illustrates the results for $\mathrm{Kn}=0.01$ obtained using mesh1 and mesh 2, with reference DSMC solutions from DS2V and SPARTA provided for comparison. The ratio of DSMC in the hybrid method varies significantly across the flow field, ranging from 0.1 to 1.0. As the mesh resolution increases from the far field to the pre-shock region, the ratio of DSMC exhibits a corresponding increase, indicating enhanced reliance on DSMC in hybrid DSMC-Shakhov method. In contrast, within the post-shock region, the strong compression effect of the bow shock leads to a sharp reduction in DSMC weight. Along the stagnation line, where local mesh refinement is implemented to resolve the boundary layer efficiently, the ratio of DSMC approaches unity, demonstrating nearly complete dependence on the DSMC method. Given DSMC's ability to accurately resolve non-equilibrium physics, increasing its ratio in shock wave and near-wall regions could substantially enhance the accuracy of flow field predictions. Fig.\ref{Ma5Kn001-zdx-temp} presents the temperature distribution along the stagnation line, where the hybrid method matches well with the DSMC results from DS2V and SPARTA. The surface pressure, shear stress, and heat flux on the cylinder wall are shown in Fig.\ref{Cylinder-Ma5Kn001-surface}, where simulation results using both mesh1 and mesh2 exhibit good consistency with benchmark data. However, the stagnation line temperature predictions from mesh2 are better than those of mesh1, which indicates that mesh resolution near the shock wave significantly impacts stagnation line accuracy while having negligible effect on numerical results near the wall.

\begin{figure}
		\centering
		\subfigure[]{\label{Ma5Kn01-pres}\includegraphics[width=0.45\textwidth]{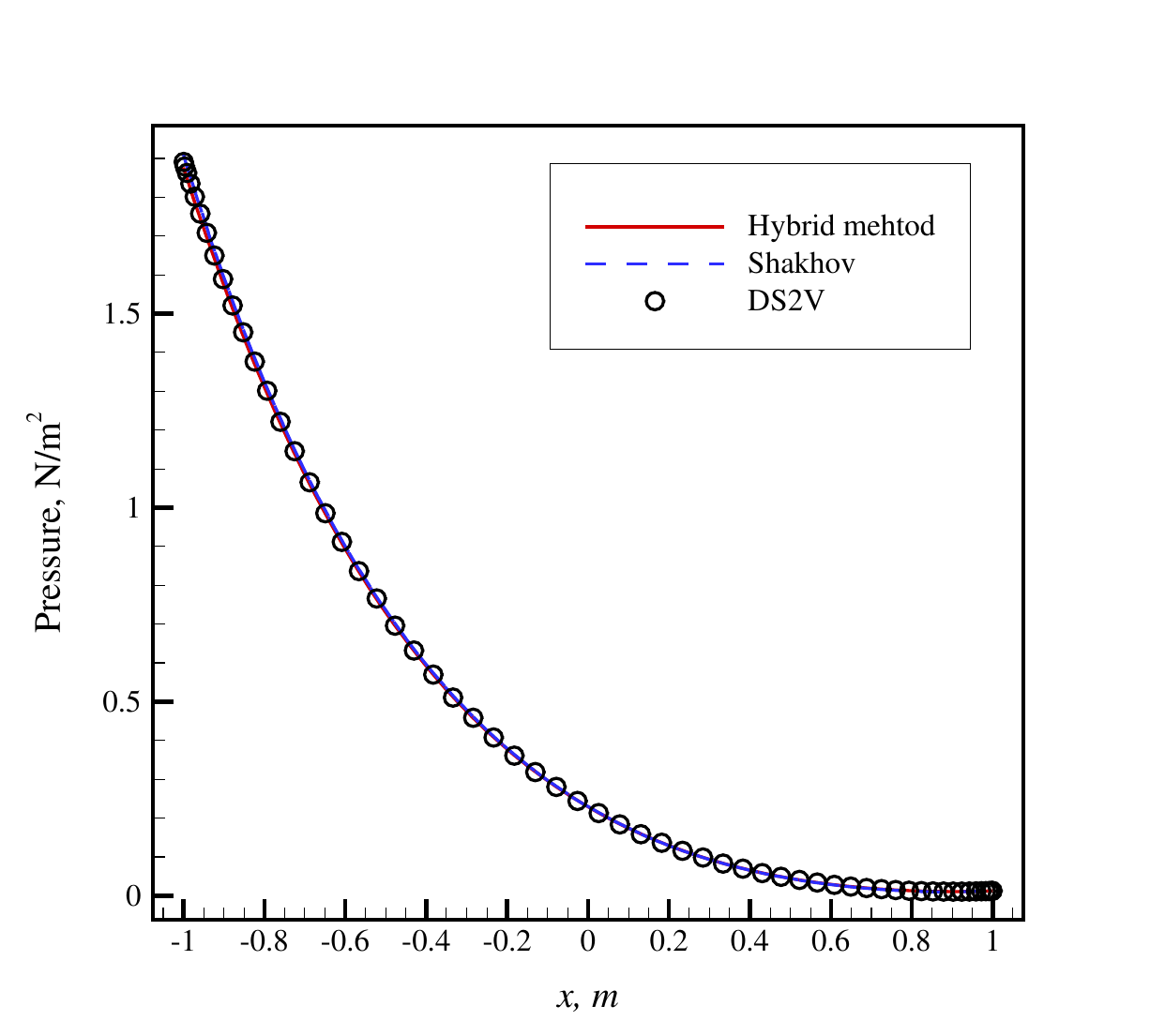}}
		\subfigure[]{\label{Ma5Kn01-tau}\includegraphics[width=0.45\textwidth]{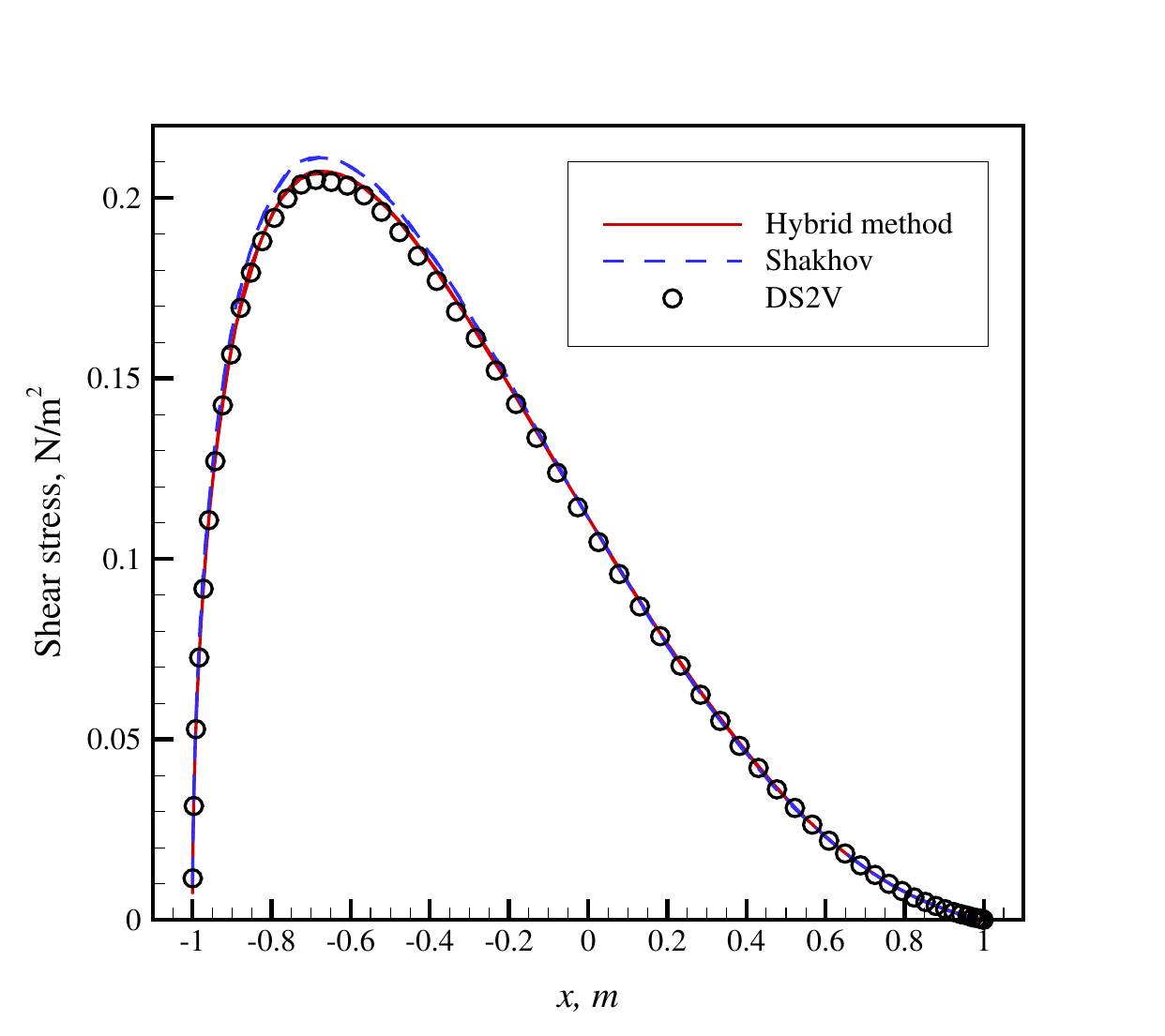}}
		\subfigure[]{\label{Ma5Kn01-q}\includegraphics[width=0.45\textwidth]{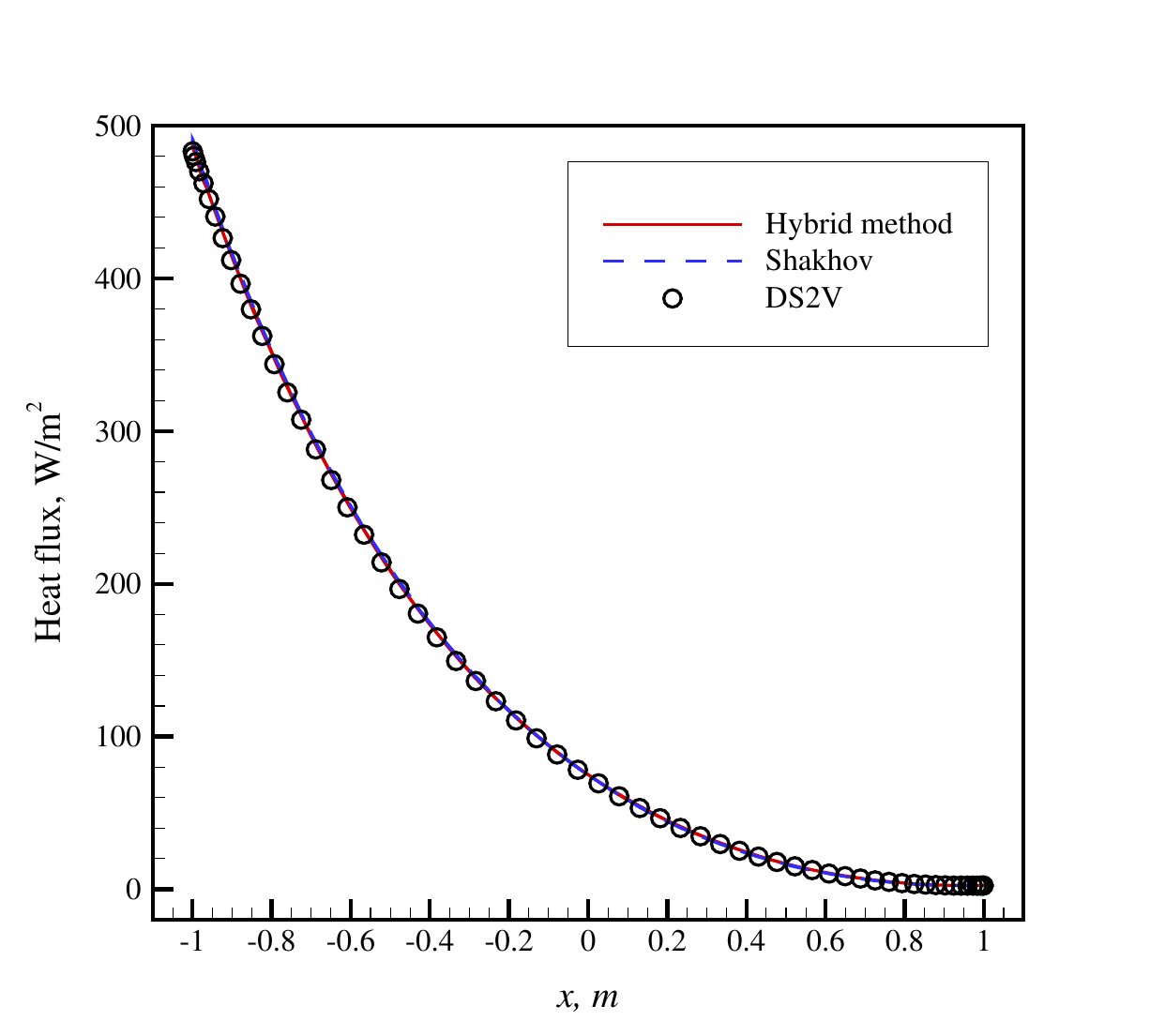}}
		\caption{\label{Cylinder-Ma5Kn01-surface} Surface results of hypersonic cylinder flow for $\mathrm{Ma}=5$, $\mathrm{Kn}=0.1$: (a) pressure, (b) shear stress, (c) heat flux. Solid line: hybrid DSMC-Shakhov; Dashed line: SP-Shakhov method, Circle symbol: DSMC from DS2V.}
	\end{figure}

	The flow field results for the $\mathrm{Kn}=0.1$ case are presented in Fig.\ref{Cylinder-Ma5Kn01-flow}, with comparisons to SP Shakhov and DS2V-based DSMC solutions. Here, the DSMC weighting ratio is unity throughout the domain, indicating full equivalence between the hybrid method and pure DSMC method. Fig.\ref{Ma5Kn01-zdx} shows the temperature distribution along the stagnation line, where the hybrid method achieves excellent agreement with DS2V reference data. The pressure, shear stress and heat flux on the wall of cylinder are shown in Fig.\ref{Cylinder-Ma5Kn01-surface}. All surface results predicted by hybrid method are in good agreement with those from DSMC method, with particularly improved shear stress accuracy over the SP Shakhov results, demonstrating the hybrid approach's superior predictive capability.

    \subsection{Jet expansion into the vacuum}~{}
	Jet expansion into a vacuum is another typical multi-scale flow problem, where the gas transitions from continuum to rarefied regimes during plume expansion. This configuration has been previously investigated using the conservative discrete unified gas-kinetic scheme (CDUGKS)\cite{chen2020compressible}. This study investigates a representative case with Mach number $\mathrm{Ma}=2.19$ and Knudsen number $\mathrm{Kn}=10^{-3}$, where the nozzle exit width serves as the characteristic reference length.
	
\begin{figure}
		\centering
		\includegraphics[width=0.6\textwidth]{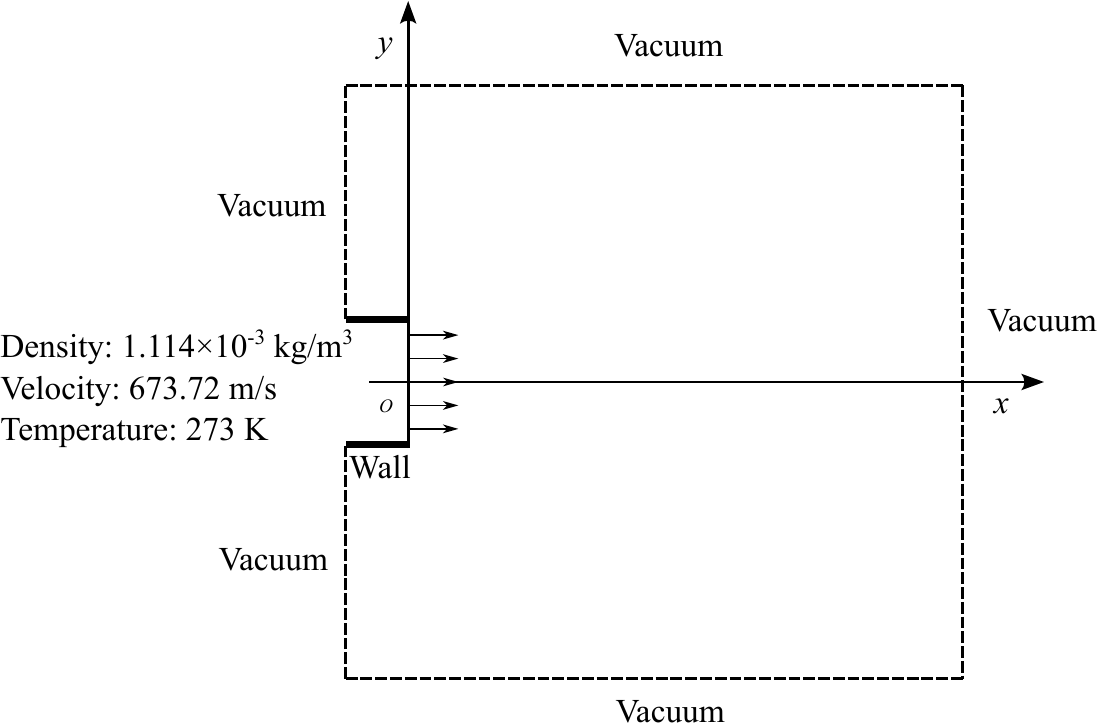}
		\caption{\label{jet_geo}Geometry of the jet flow expanding into vacuum environment.}	
	\end{figure}

\begin{figure}
		\centering
		\includegraphics[width=0.6\textwidth]{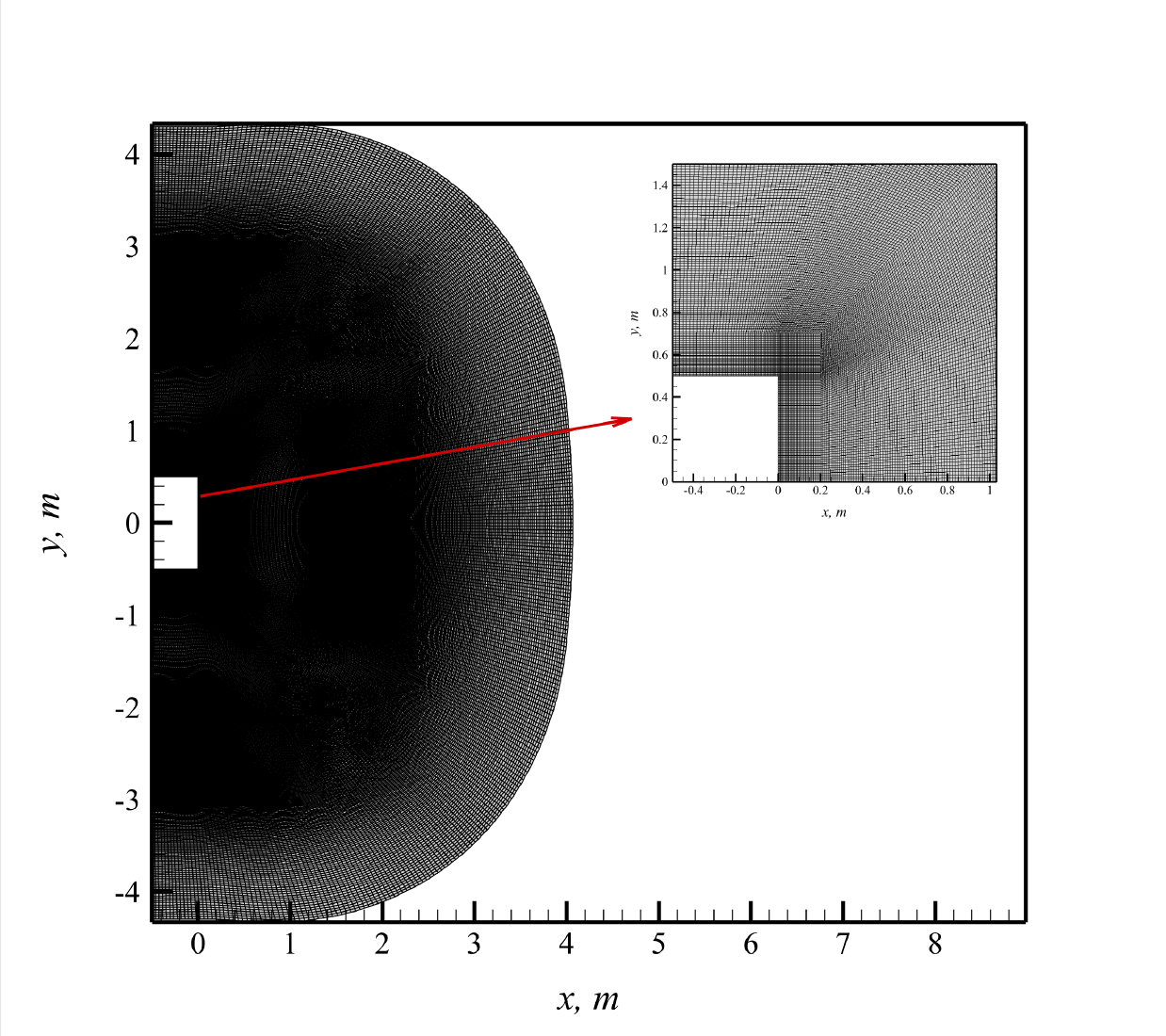}
		\caption{\label{jet-mesh} The physical space mesh (total 104912 quadrangle cells) used in jet expansion into vacuum environment.}
	\end{figure}

	The computational domain geometry of the jet flow is illustrated in Fig. \ref{jet_geo}. The width of the nozzle outlet, which is also the inlet of the jet flow, is $L=1m$. The density of jet flow is $1.114161 \times 10^{-4} kg/m^{3}$, the temperature and velocity are 273K and $1537.44 m/s$, respectively. The coordinate origin $(0,0)$ is located at the geometric center of the nozzle exit. Both upper and lower nozzle walls are maintained at a constant temperature of 273K and treated as fully diffusive reflective boundaries. All variables are nondimensionalized according to Eq. \ref{dimensionless variables}. The computational grid consists of 104192 quadrangle cells, as shown in Fig.\ref{jet-mesh}. The number of simulated particles in each cell is initially set to $10^3$.

\begin{figure}
		\centering
		\subfigure[]{\label{Jet-t=0.5}\includegraphics[width=0.45\textwidth]{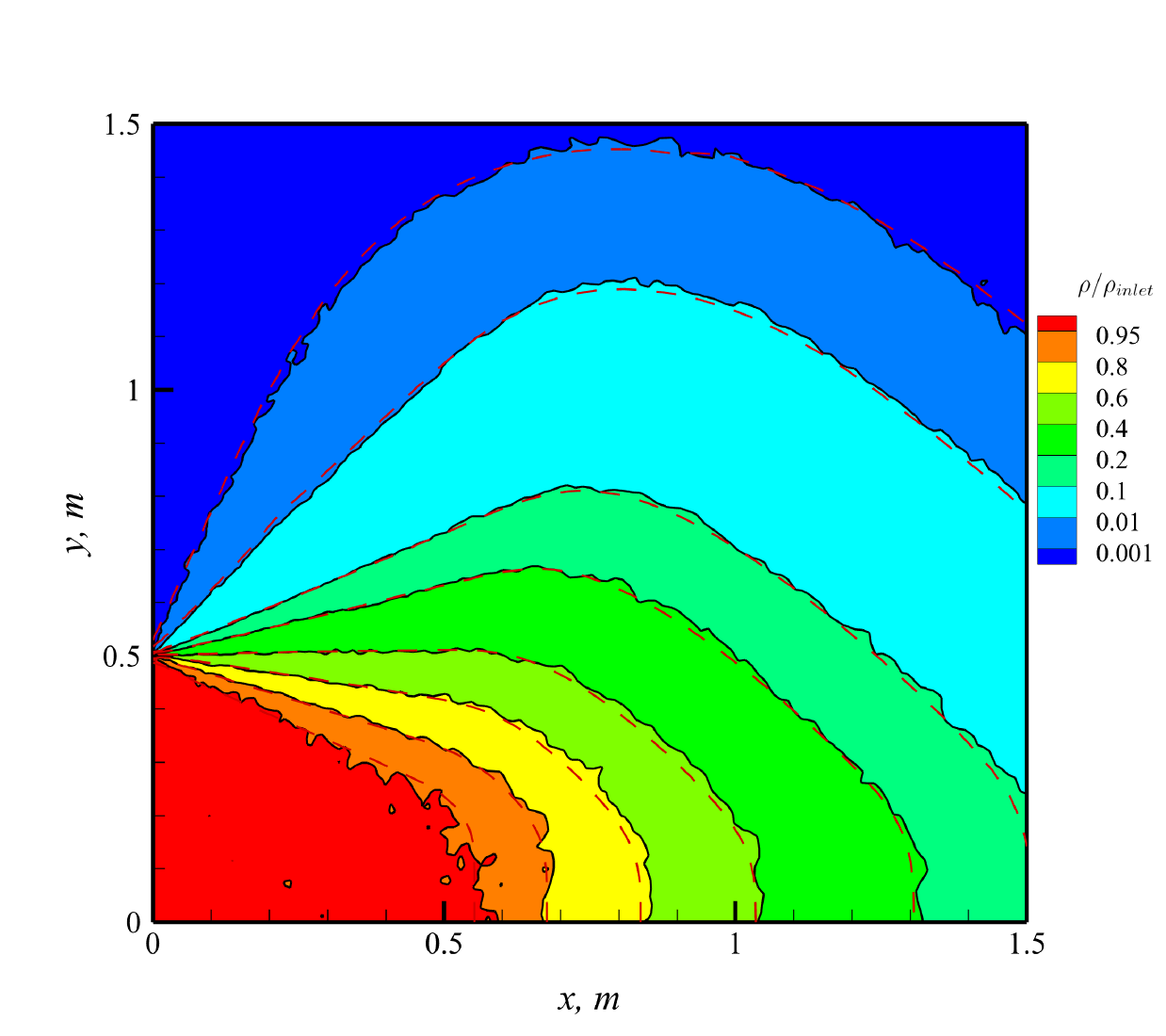}}
		\subfigure[]{\label{Jet-t=1.0}\includegraphics[width=0.45\textwidth]{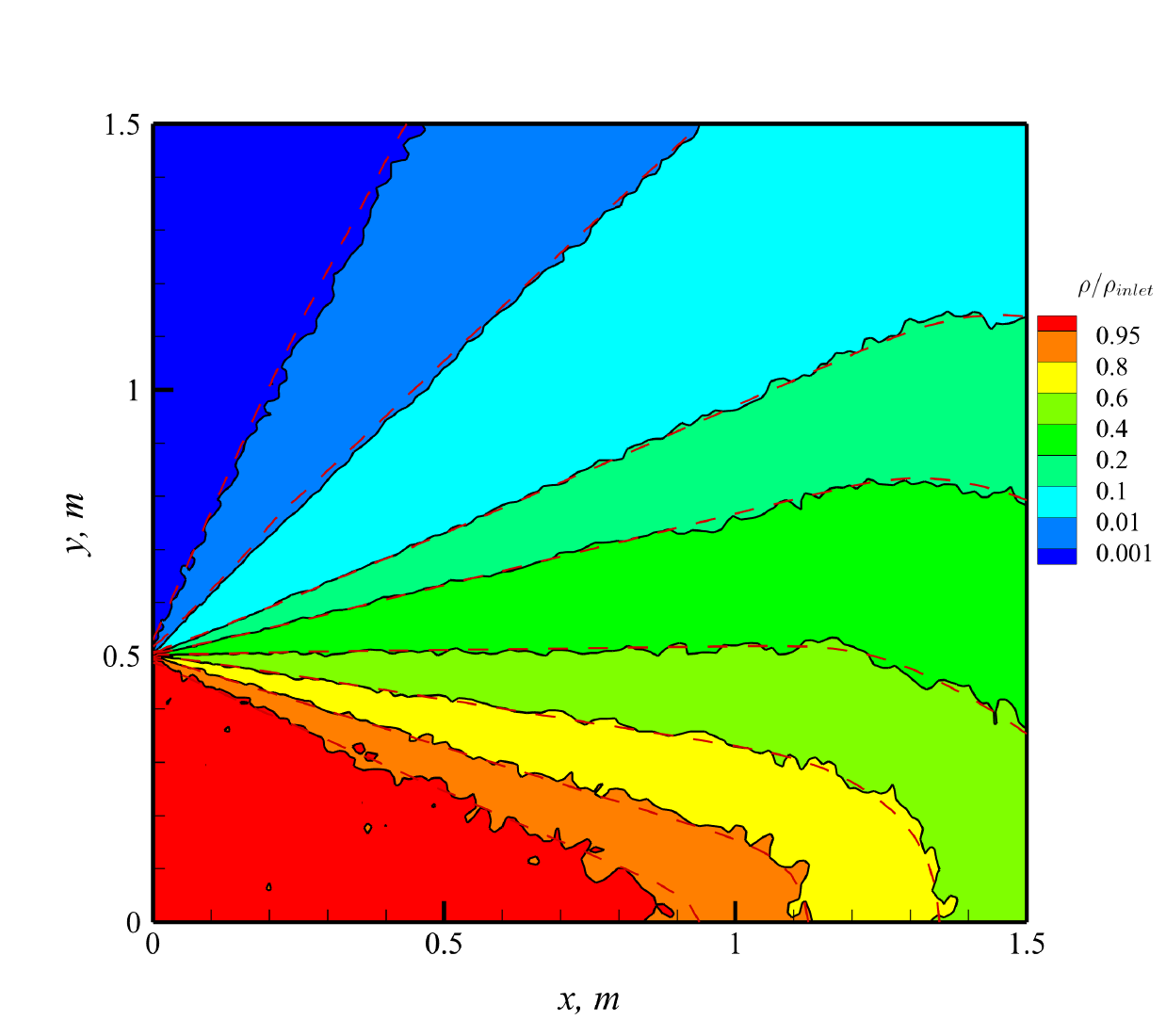}}
		\subfigure[]{\label{Jet-t=2.0}\includegraphics[width=0.45\textwidth]{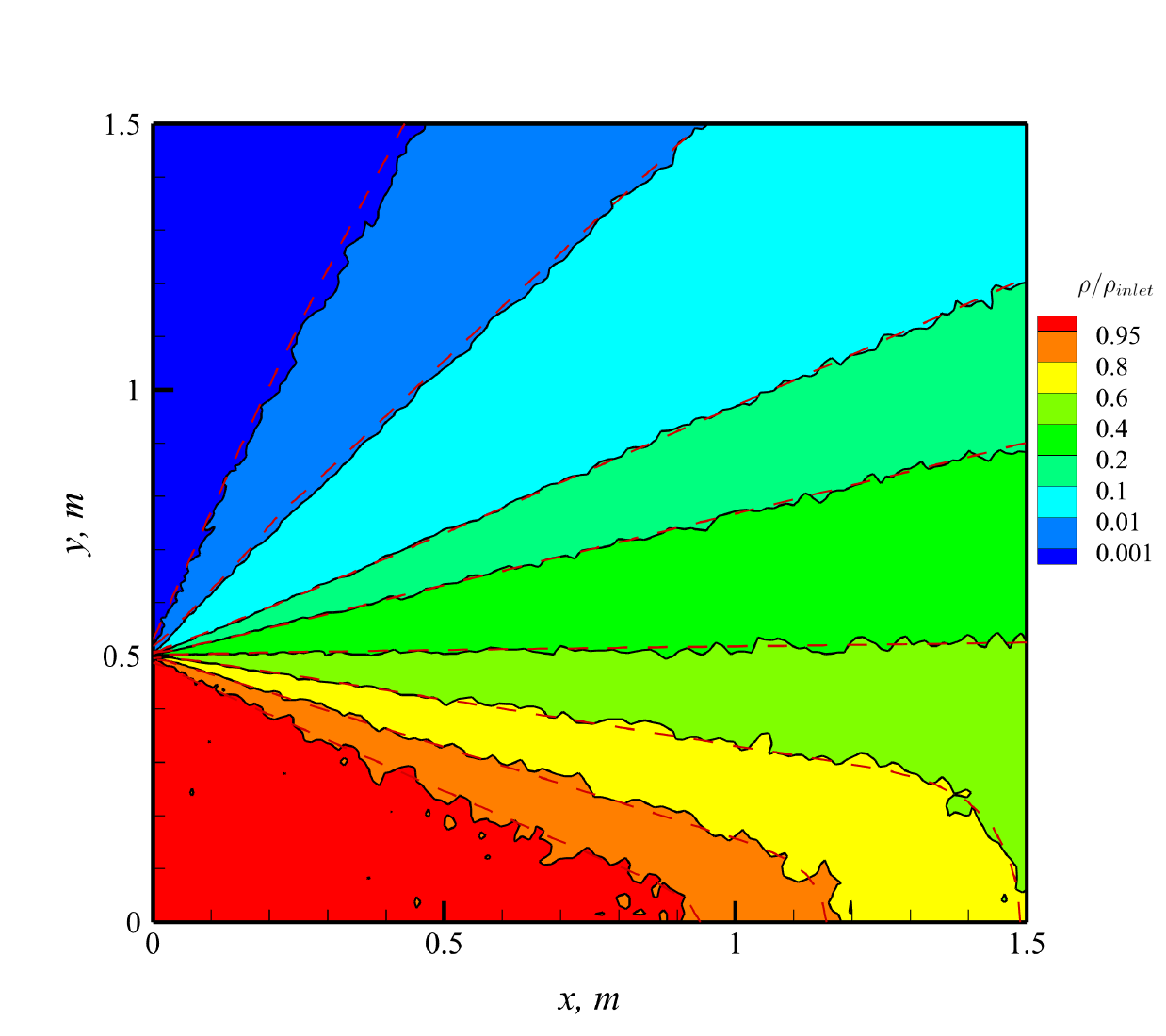}}
		\subfigure[]{\label{Jet-steady}\includegraphics[width=0.45\textwidth]{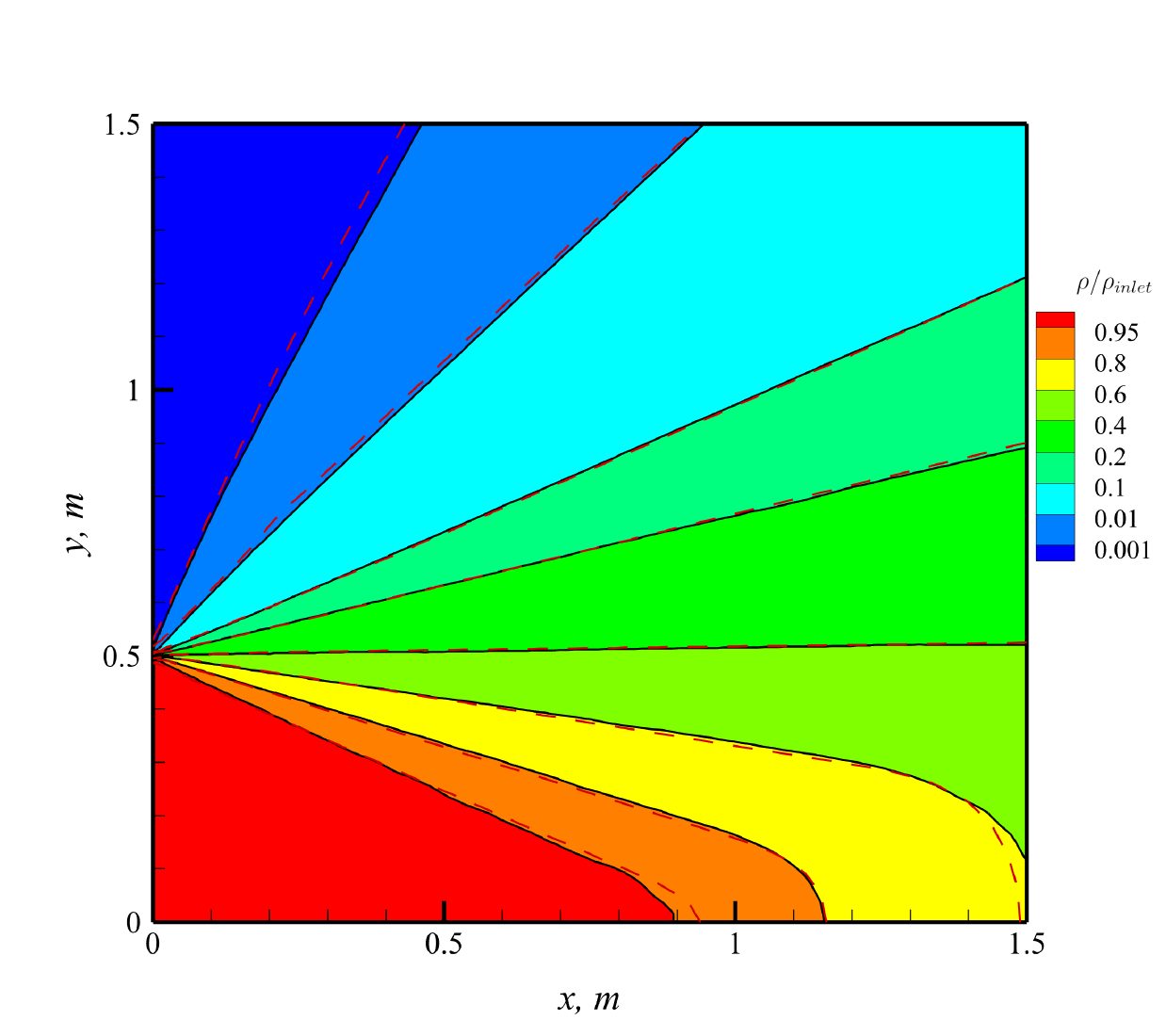}}
		\caption{\label{Jet} Density contours of jet flow for $\mathrm{Ma}=2.19$, $\mathrm{Kn}=10^{-3}$ at different time steps: (a) $t=0.5$, (b) $t=1.0$, (c) $t=2.0$, (d) steady state. Solid line: hybrid DSMC-Shakhov method; Dashed line: CDUGKS data of Ref\cite{chen2020compressible}.}
	\end{figure}

\begin{figure}
		\centering
		\subfigure[]{\label{Jet-Kn}\includegraphics[width=0.45\textwidth]{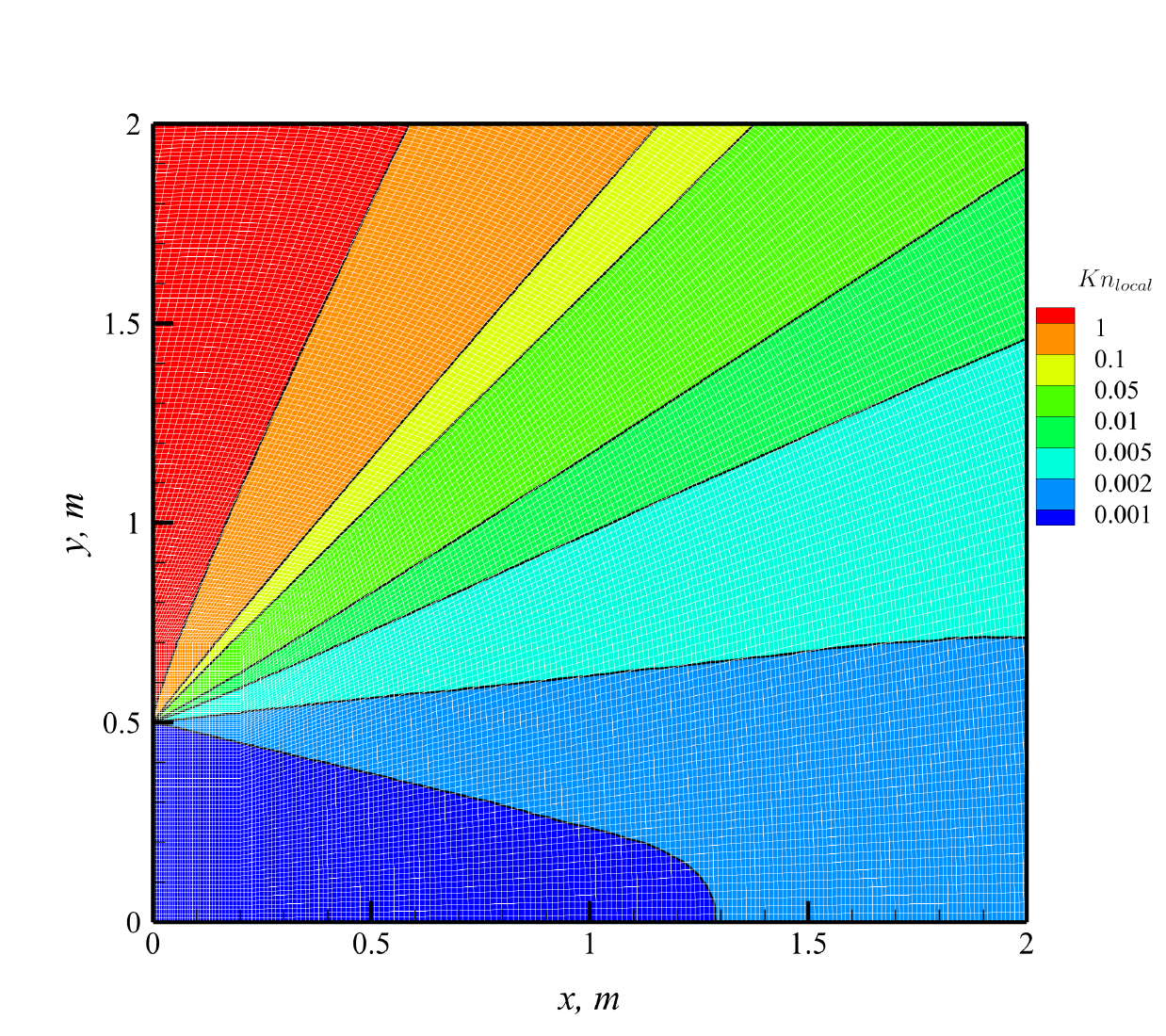}}
		\subfigure[]{\label{Jet-weight}\includegraphics[width=0.45\textwidth]{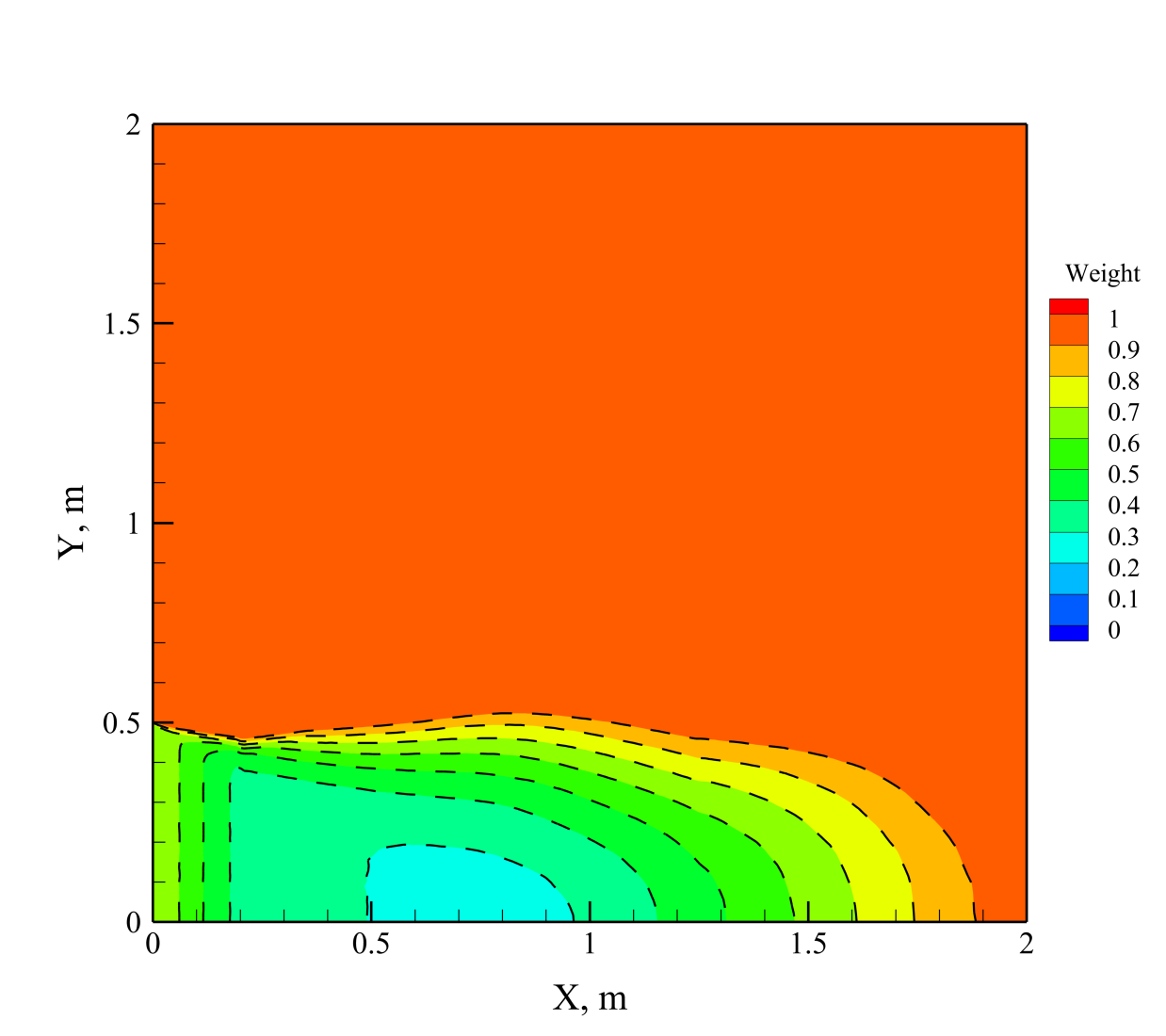}}
		\caption{\label{Jet-Kn-weight} Results of jet flow for $\mathrm{Ma}=2.19$, $\mathrm{Kn}=10^{-3}$ at steady state: (a) local Knudsen number field with computational mesh (white gridlines), (b) ratio of DSMC field.}
	\end{figure}

	The numerical results for the $\mathrm{Kn}=10^{-3}$ case at different time steps are presented in Fig.\ref{Jet}, with CDUGKS results for comparison. The density field at $t=0.5$, $t=1.0$ and $t=2.0$ are shown in Fig.\ref{Jet-t=0.5}, Fig.\ref{Jet-t=1.0} and Fig.\ref{Jet-t=2.0}, respectively, where the jet flow expands into the vacuum environment and the density decreases rapidly. The density field at steady state is shown in Fig.\ref{Jet-steady}, where the smooth contours are achieved through extensive ensemble averaging implemented in the hybrid method. These flow field results from the present hybrid method exhibit excellent agreement with the benchmark CDUGKS results. The steady-state local Knudsen number field and ratio of DSMC are present in Fig.\ref{Jet-Kn} and Fig.\ref{Jet-weight}, respectively. The ratio of SP Shakhov remains relatively low in the flow field, and the primarily contributing factor is that the computational mesh size exhibits relatively minimal variation when compared to the substantial scale (local Knudsen number) changes observed in the flow.
	
	\subsection{Hypersonic flow around a sphere}
\begin{figure}
		\centering
		\includegraphics[width=0.5\textwidth]{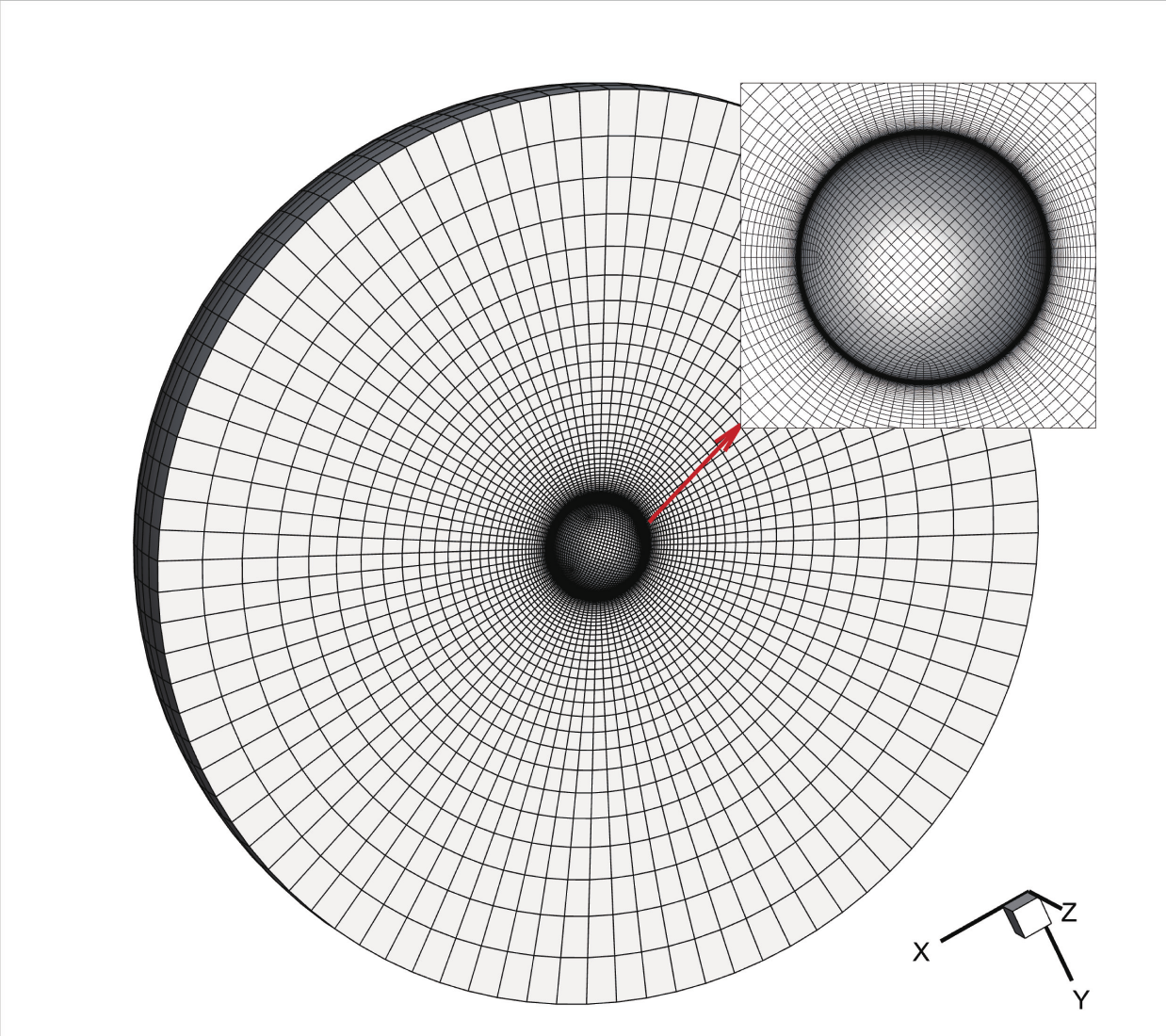}
		\caption{\label{Sphere-mesh} The section views of physical space mesh (total 197248 hexahedral cells) for the hypersonic flow over a sphere.}
	\end{figure}

~{}
	The hypersonic flow around a sphere is a classical three-dimensional multiscale flow problem. The computational setup adopted in this study follows the methodology presented in Ref. \cite{zhang2022unified}. The flow conditions are characterized by a Mach number of 10 and a Knudsen number of 0.01, with the sphere radius selected as the reference length. The characteristic length of the sphere is set to $1m$. The freestream conditions are specified as follows: a density of $5.5 \times 10^{-6} kg/m^{3}$, a temperature of 65K, and a velocity of $1501.11 m/s$. A fully diffusive wall boundary condition is applied, with the surface temperature of the sphere maintained at 302 K. The computational domain of physical space has total 197248 hexahedral cells, and a body mesh consiting of 3082 quadrilateral cells is used to define the wall surface, as shown in Fig.\ref{Sphere-mesh}.

\begin{figure}
		\centering
		\subfigure[]{\label{Sphere-temp}\includegraphics[width=0.45\textwidth]{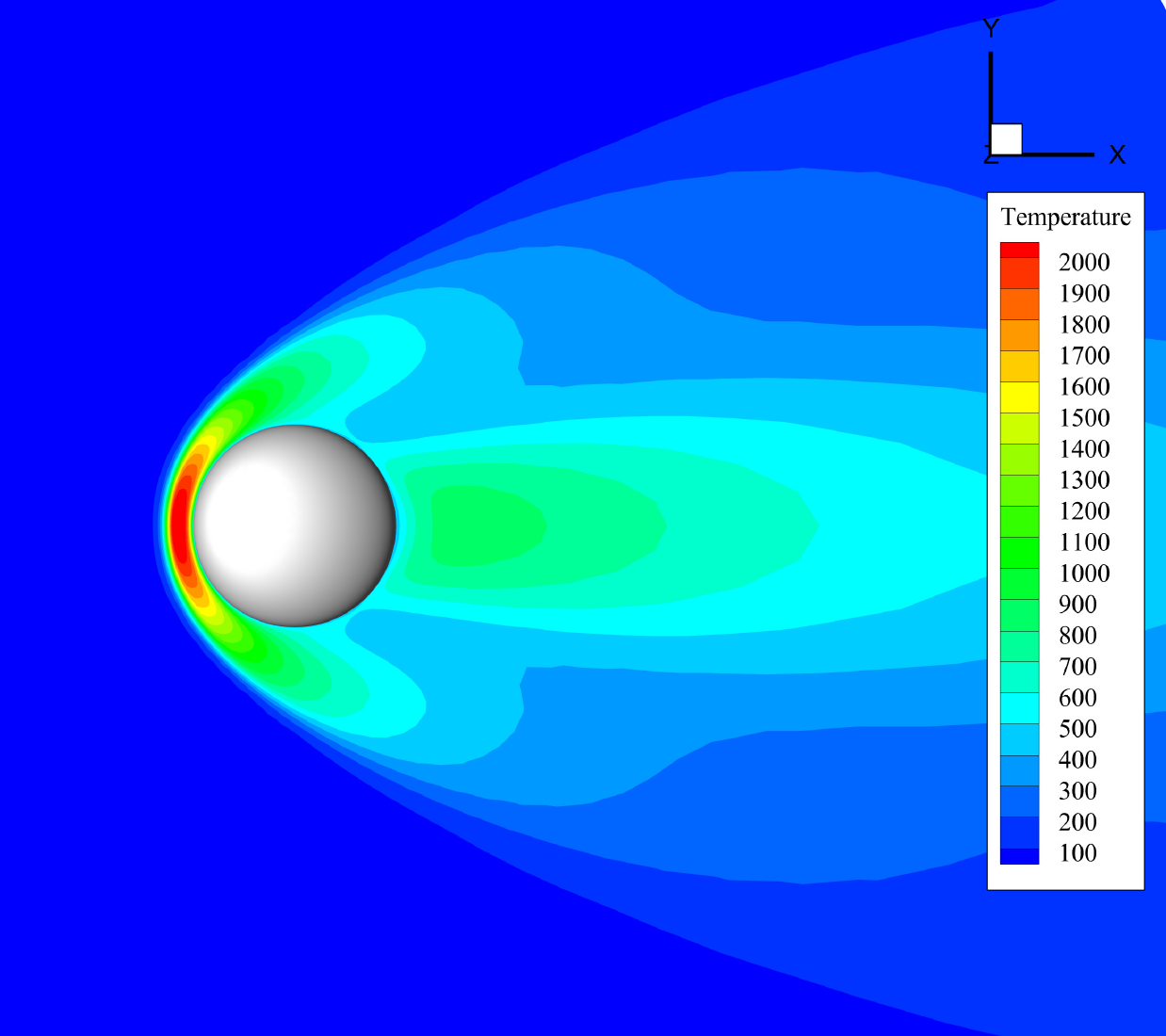}}
		\subfigure[]{\label{Sphere-Ma}\includegraphics[width=0.45\textwidth]{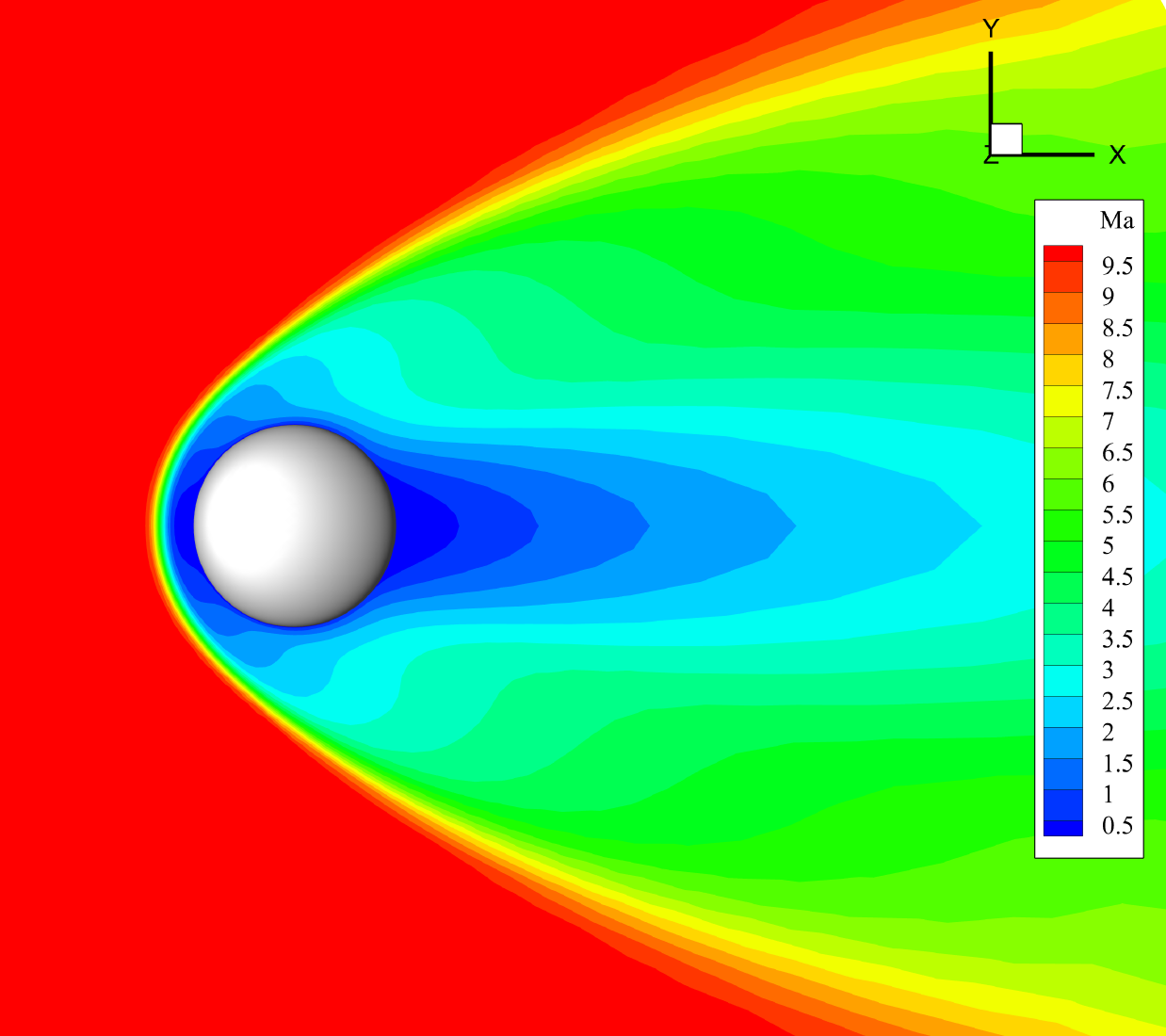}}
		\subfigure[]{\label{Sphere-weight}\includegraphics[width=0.45\textwidth]{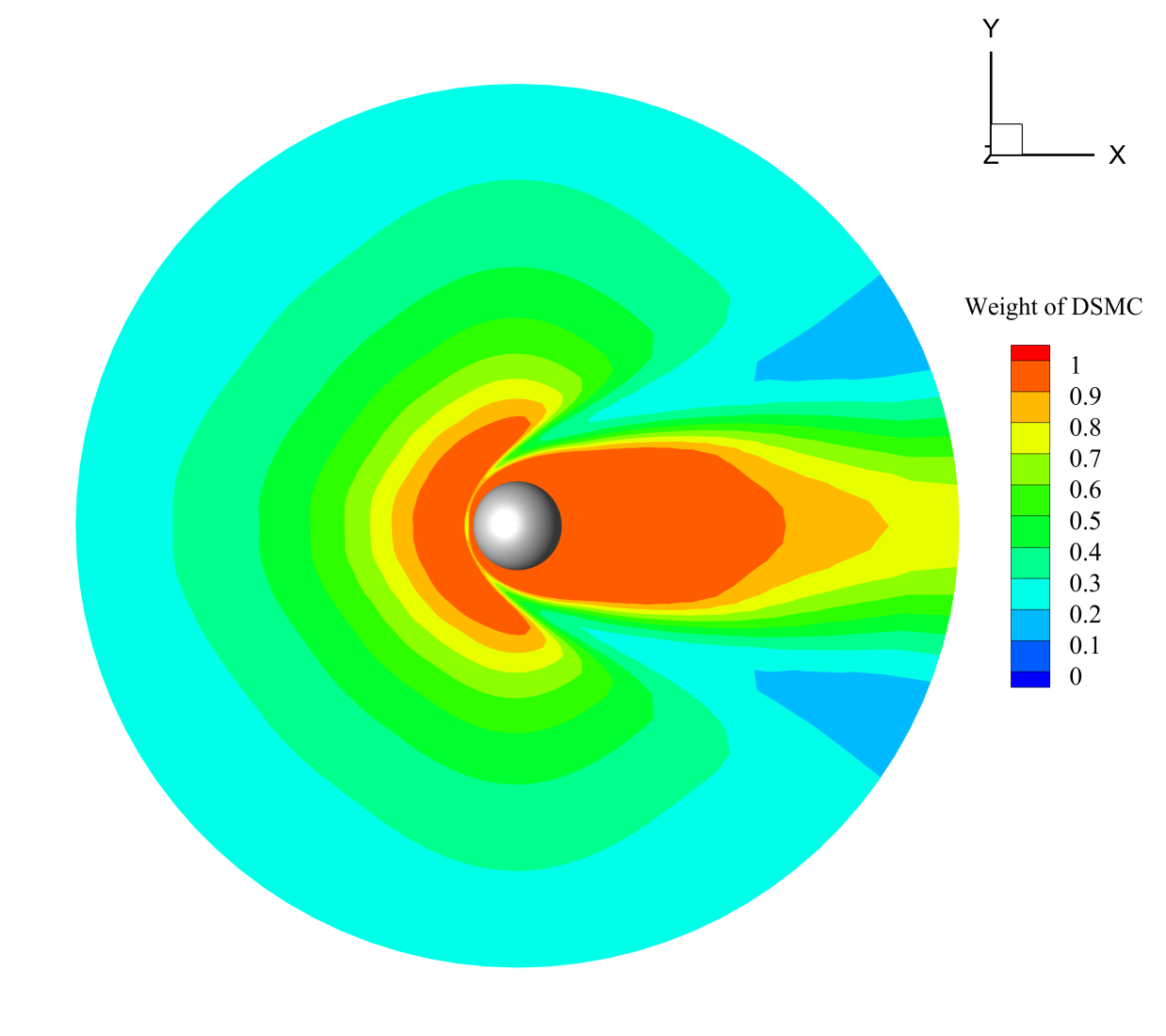}}
		\caption{\label{Sphere contour} Contours of the sphere flow at slice $Z=0$ for $\mathrm{Ma}=10$, $\mathrm{Kn}=0.01$: (a) temperature field, (b) Mach number field, (c) ratio of DSMC.}
	\end{figure}

\begin{figure}
		\centering
		\subfigure[]{\label{Sphere-pres}\includegraphics[width=0.45\textwidth]{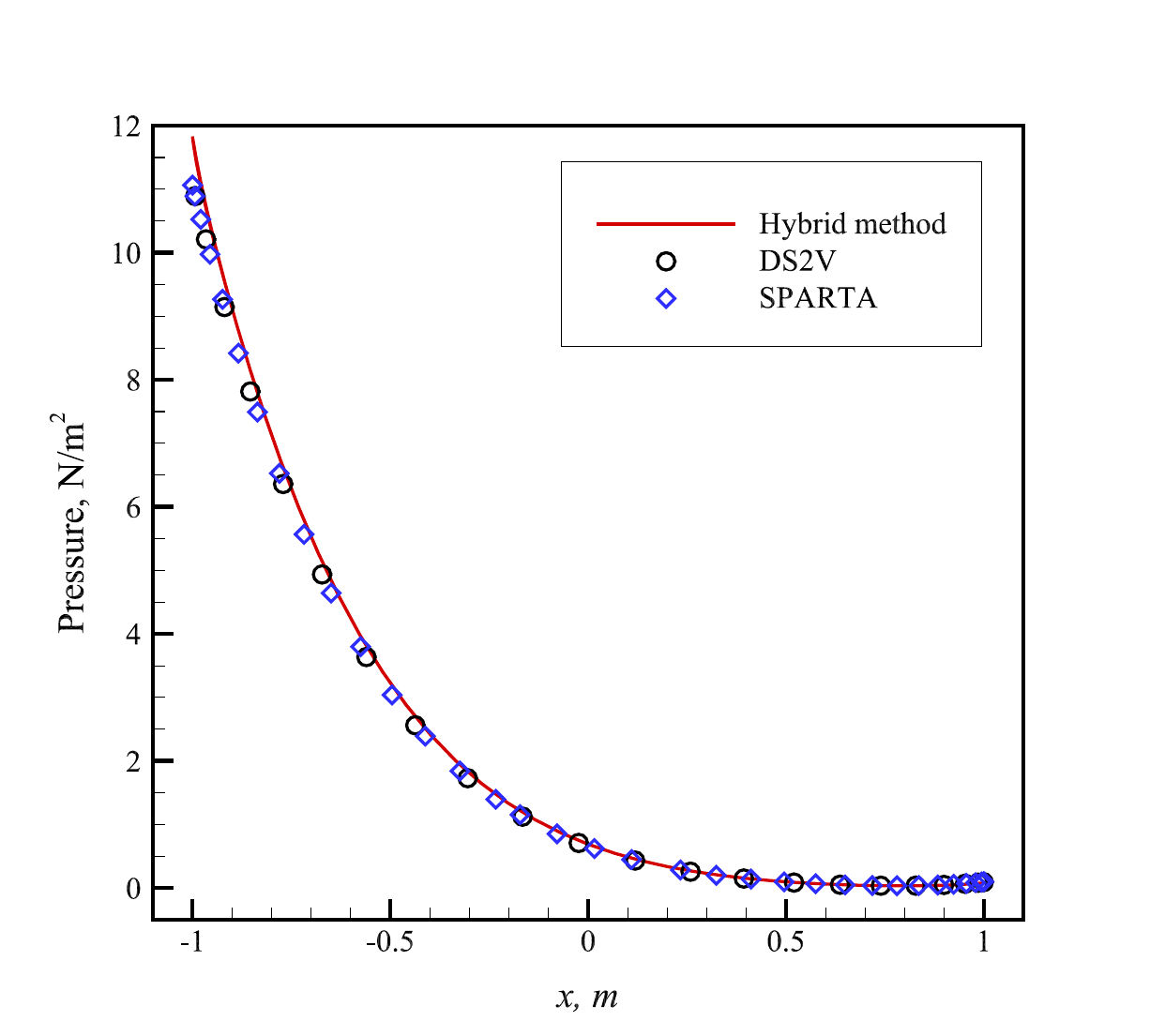}}
		\subfigure[]{\label{Sphere-tau}\includegraphics[width=0.45\textwidth]{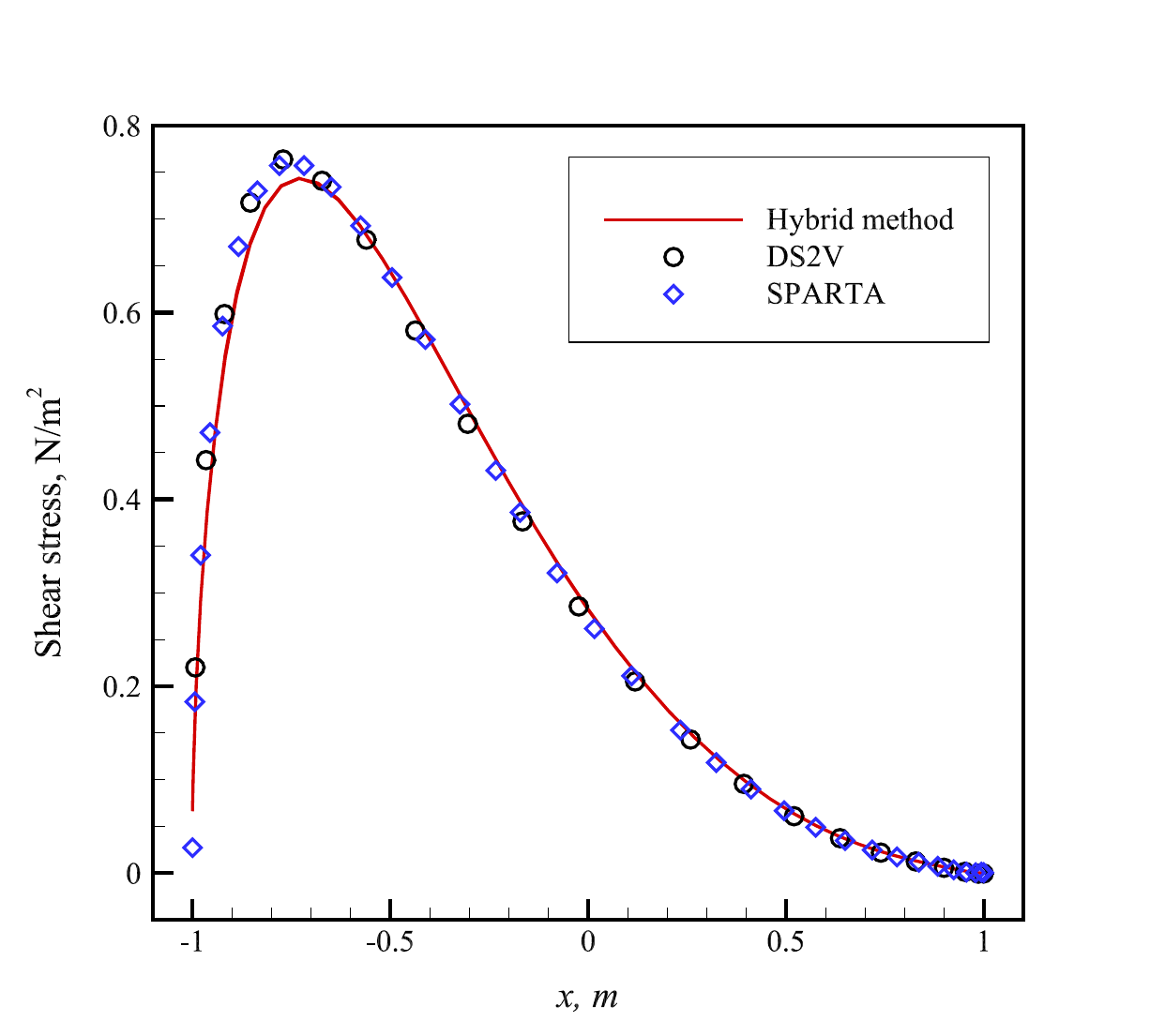}}
		\subfigure[]{\label{Sphere-q}\includegraphics[width=0.45\textwidth]{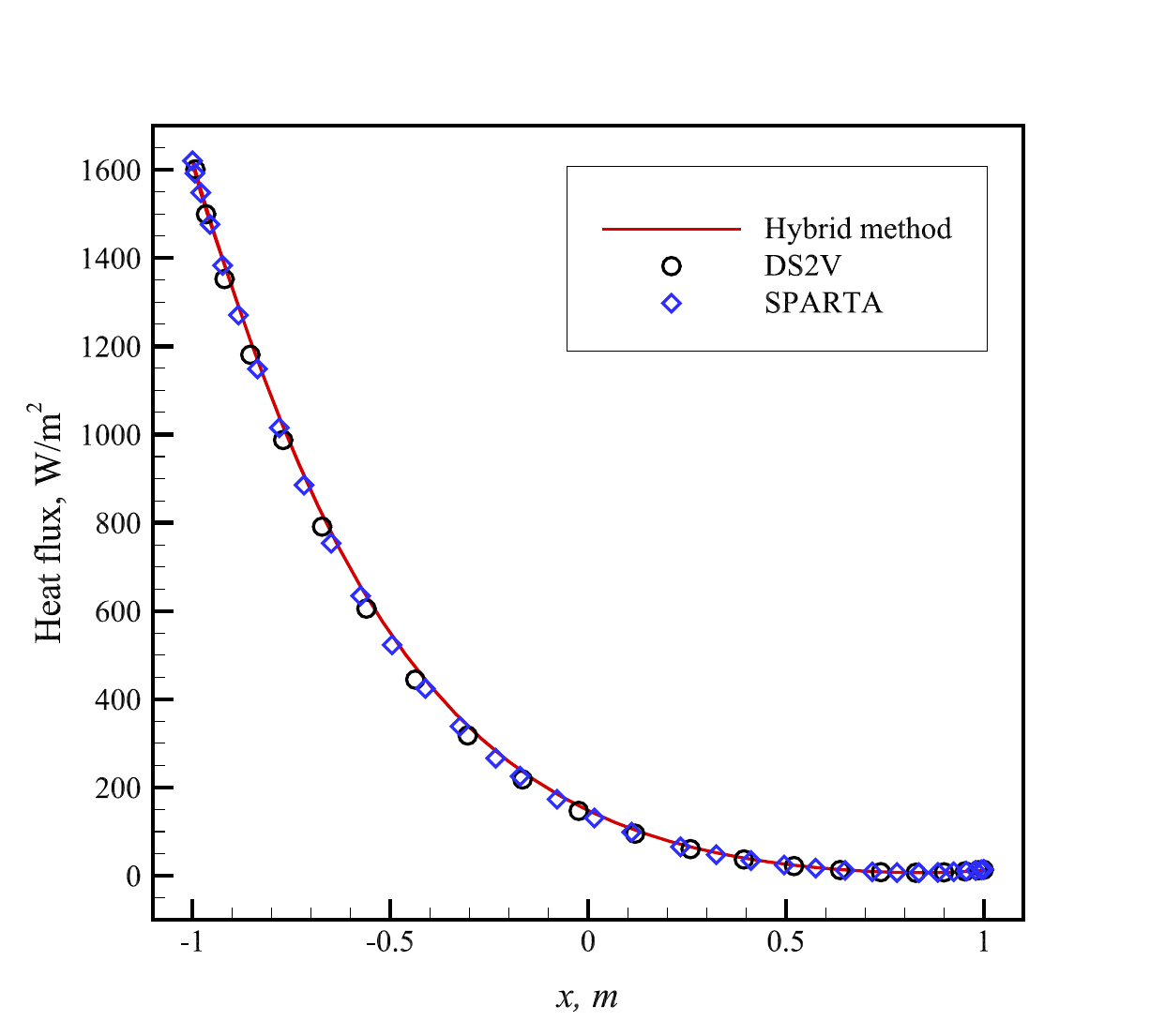}}
		\caption{\label{Sphere coefficient} Surface results of the sphere flow at slice $Z=0$ for $\mathrm{Ma}=10$, $\mathrm{Kn}=0.01$: (a) pressure, (b) shear stress, (c) heat flux. Solid line: hybrid DSMC-Shakhov; Circle symbol: DSMC from DS2V\cite{zhang2022unified}, Diamond symbol: DSMC from SPARTA.}
	\end{figure}

	The temperature and Mach number contours at slice $Z=0$ are illustrated in Fig.\ref{Sphere-temp} and Fig.\ref{Sphere-Ma}, respectively. Fig.\ref{Sphere-weight} presents the ratio of DSMC at slice $Z=0$. Similar to the 2-D hypersonic flow past a cylinder, the ratio of DSMC exhibits significant variations across the flow field. The maximum values of ratio of DSMC occur in the pre-shock and near-wall regions, approaching unity. The surface results of pressure, shear stress and heat flux at slice $Z=0$ are depicted in Fig.\ref{Sphere coefficient}.  The hybrid method predictions show excellent agreement with both DS2V and SPARTA data, further validating the accuracy of the proposed hybrid Shakhov-DSMC method.
	
	\subsection{Hypersonic flow around X-38 like vehicle}
\begin{figure}
		\centering
		\subfigure[]{\label{X38-mesh}\includegraphics[width=0.45\textwidth]{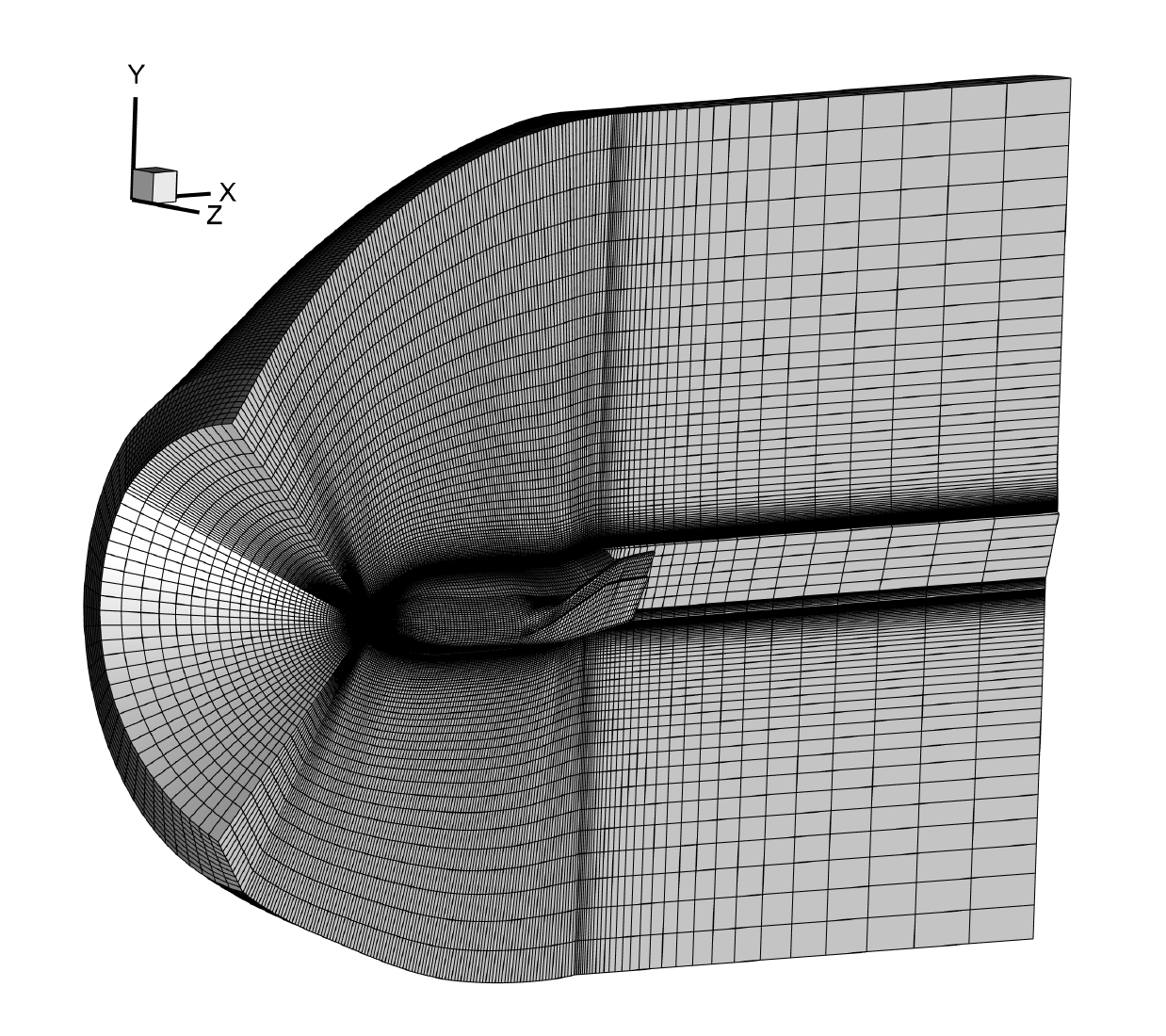}}
		\subfigure[]{\label{X38-mesh-surf}\includegraphics[width=0.45\textwidth]{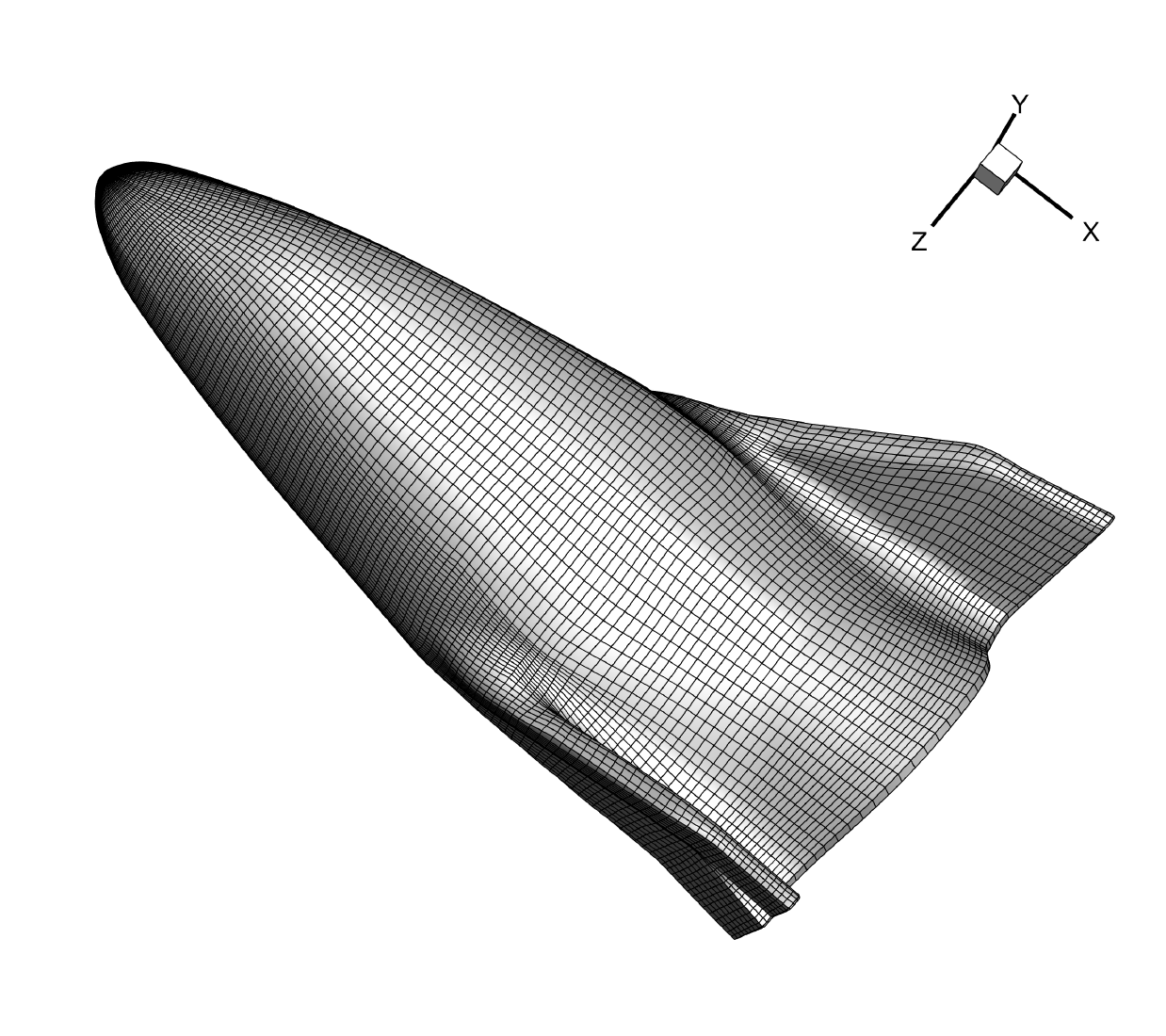}}
		\caption{\label{X38 mesh} The physical space mesh adopted in numerical simulation for hypersonic flow over an X38-like space vehicle. (a) The section view of the mesh, total 961080 hexahedral cells, (b) meshes on the wall surface.}
	\end{figure}

~{}
	The hypersonic flow around an X-38 like configuration is investigated to demonstrate the capability of the hybrid method in simulating three-dimensional hypersonic flows over complex geometries. This case was previously studied by Jiang et al.\cite{jiang2019implicit} using UGKS method, and one representative condition is selected for numerical calculation in this work. The freestream conditions are set at $\mathrm{Ma}=8$ and $\mathrm{Kn}=1.68\mathrm{E}-3$, with a reference length of $0.28 m$. The angle of attack is set to $20^{\circ}$. The freestream number density is $1.68392 \times 10^{18} m^{-3}$, with a temperature of 56K and a velocity vector of $(1047.33, 381.233, 0)m/s$. A fully diffusive wall boundary condition is applied, with the surface temperature maintained at 300 K. The reference area $A_{ref}$ for calculating aerodynamic coefficients is $2.41\times 10^{-2} m^2$. The computational domain is discretized into 961080 hexahedral cells, as shown in Fig.\ref{X38 mesh}.

\begin{figure}
		\centering
		\subfigure[]{\label{X38-temp}\includegraphics[width=0.45\textwidth]{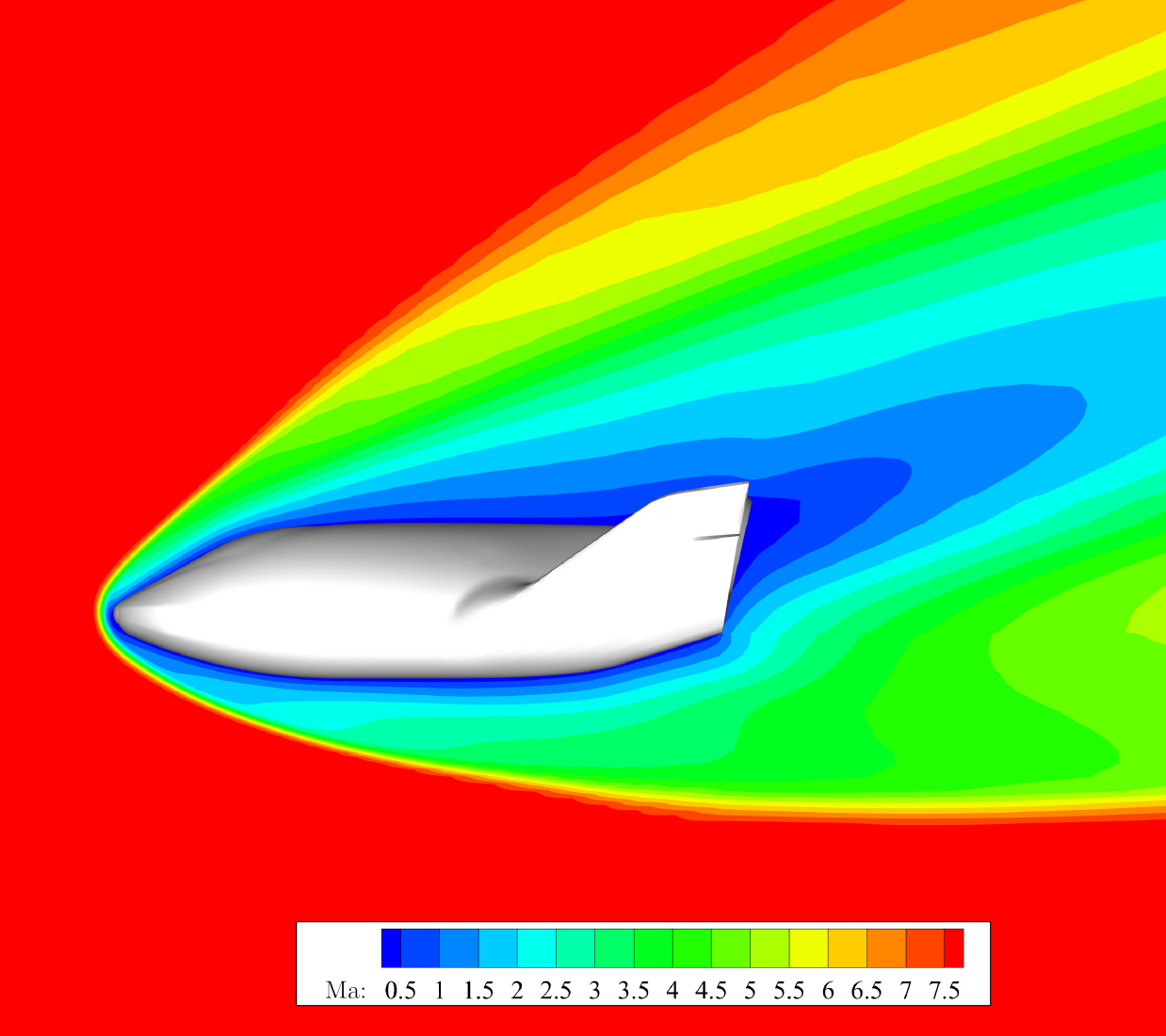}}
		\subfigure[]{\label{X38-Cp}\includegraphics[width=0.45\textwidth]{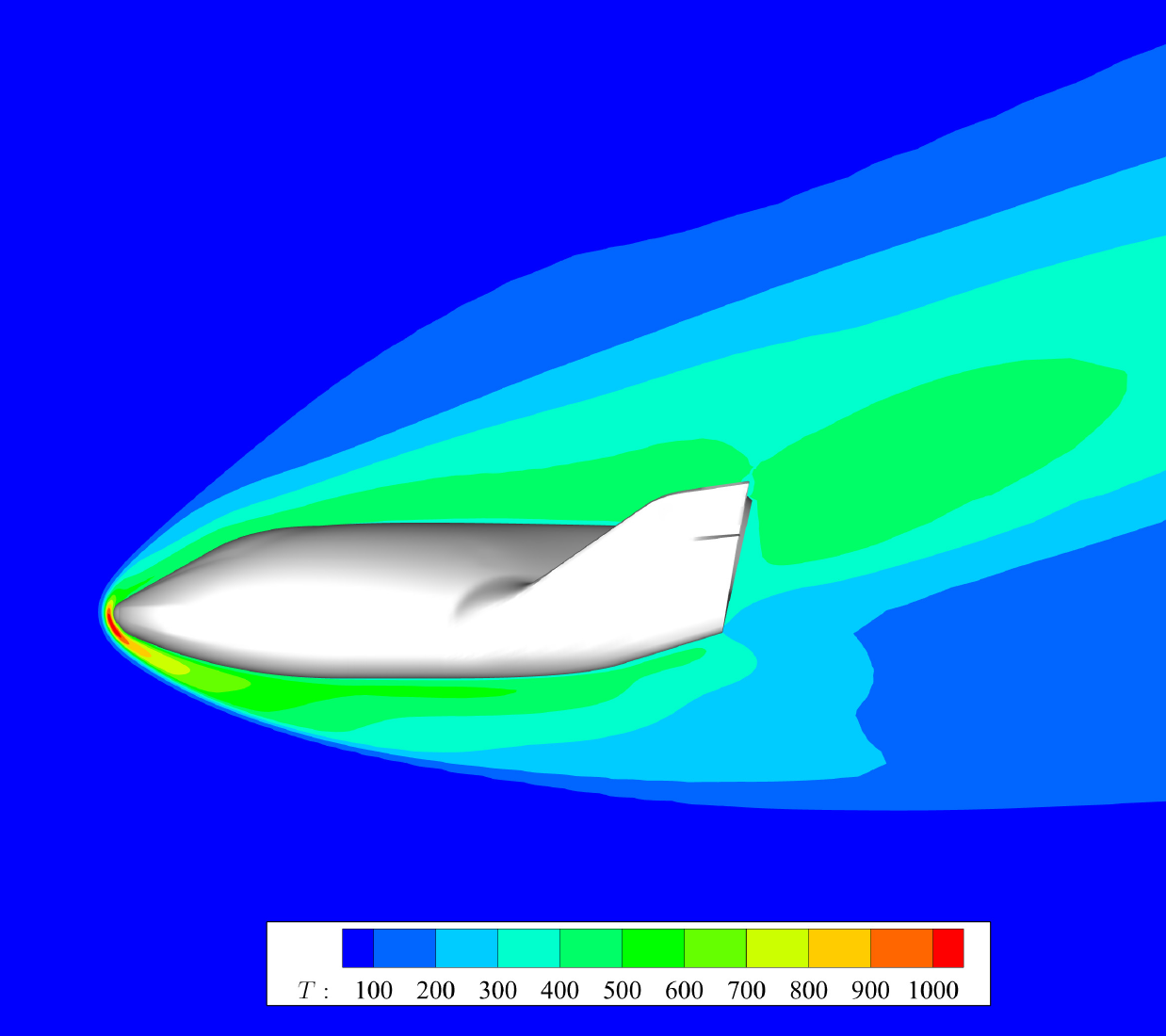}}
		\caption{\label{X38 contour-flow} Flow field results of the X-38 like vehicle for $\mathrm{Ma}=8$, $\mathrm{Kn}=1.68\mathrm{E}-3$: (a) temperature contour at slice $Z=0$, (b) Mach number contour at slice $Z=0$.}
	\end{figure}

\begin{figure}
		\centering
		\subfigure[]{\label{X38-temp-2}\includegraphics[width=0.45\textwidth]{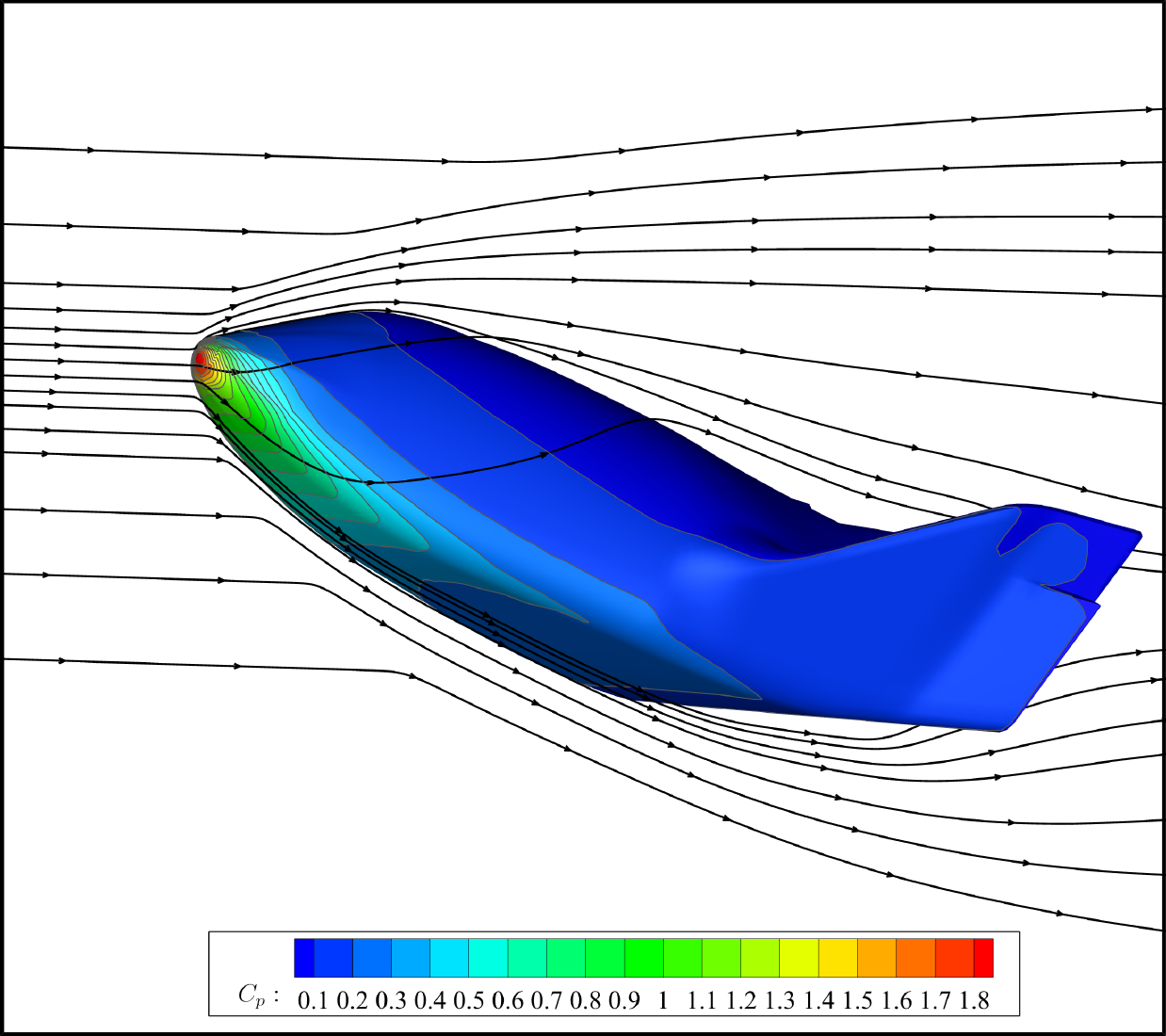}}
		\subfigure[]{\label{X38-Cp-2}\includegraphics[width=0.45\textwidth]{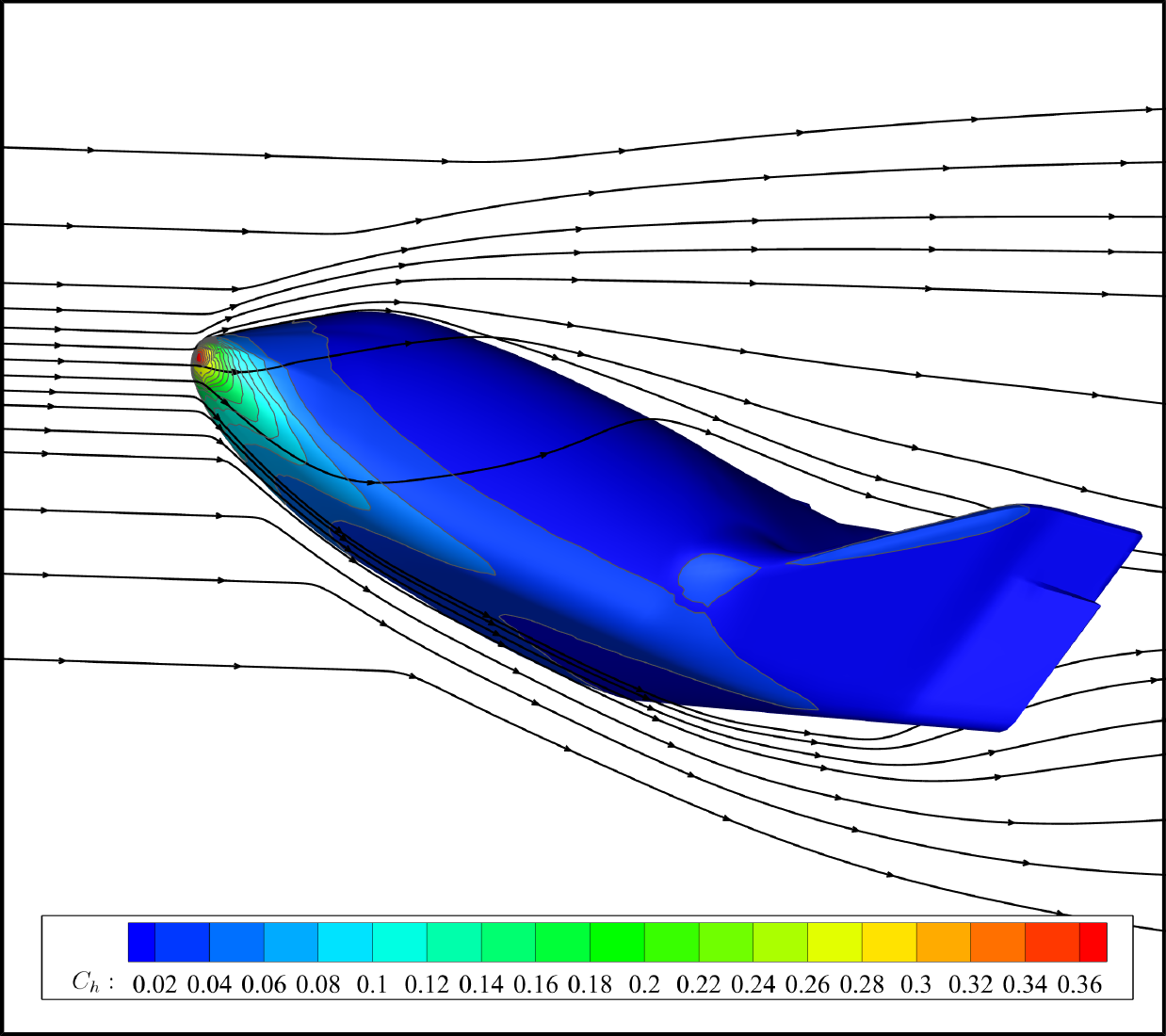}}
		\caption{\label{X38 contour-surf} Surface results of the X-38 like vehicle for $\mathrm{Ma}=8$, $\mathrm{Kn}=1.68\mathrm{E}-3$: (a) surface pressure coefficient with spatial streamlines, (b) surface heat flux coefficient with spatial streamlines.}
	\end{figure}

\begin{figure}
		\centering
		\includegraphics[width=0.9\textwidth]{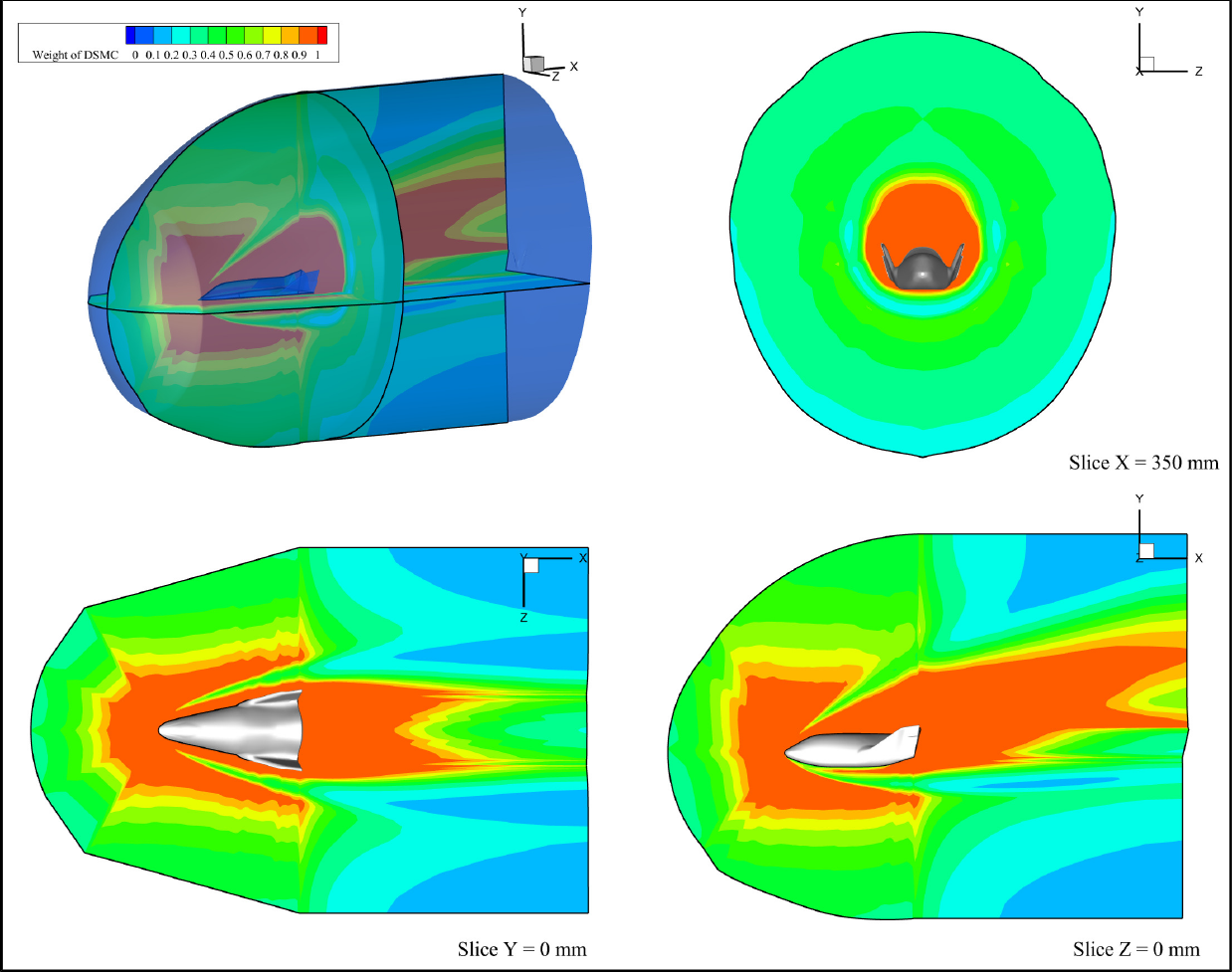}
		\caption{\label{X38-weight} Ratio of DSMC around the X-38 like vehicle for $\mathrm{Ma}=8$, $\mathrm{Kn}=1.68\mathrm{E}-3$.}
	\end{figure}

	Fig.\ref{X38 contour-flow} presents the temperature and Mach number contours at slice $Z=0$. The surface pressure coefficient and heat flux coefficient distributions with spatial streamlines are illustrated in Fig.\ref{X38 contour-surf}. The dimensionless surface coefficients are normalized as follows
	\begin{equation}
		\begin{aligned}
			& C_{p}=\frac{{p_s}-{{p}_{\infty}}}{\left( 1/2 \right){{\rho }_{\infty }}U_{^{\infty }}^{2}}, \\
			& C_{h}=\frac{h_s}{\left( 1/2 \right){{\rho }_{\infty }}U_{^{\infty }}^{3}}, \\
		\end{aligned}
	\end{equation}
	where $p_s$ is the surface pressure, $p_{\infty}$ is the pressure in freestream flow. The ratio of DSMC in the flow field is shown in Fig.\ref{X38-weight}. Consistent with observations from both the 2-D cylinder and 3-D sphere cases, the ratio of DSMC exhibits considerable spatial variation. In the pre-shock and near-wall regions, the ratio of DSMC approaches unity, which is reasonable because the DSMC method is more accurate. However, in the pre-shock region, the DSMC contribution is observed to be significantly enhanced, which can be attributed to the complex geometry of the X-38 like vehicle and the consequent variations in local mesh resolution. Table \ref{X-38 coeff} provides a quantitative comparison of aerodynamic coefficients with reference solutions from UGKS and DSMC (from RariHV). The dimensionless aerodynamic coefficients are normalized as follows
	\begin{equation}
		\begin{aligned}
			& C_{L}=\frac{L}{\left( 1/2 \right){{\rho }_{\infty }}U_{^{\infty }}^{2}{{A}_{ref}}}, \\
			& C_{d}=\frac{D}{\left( 1/2 \right){{\rho }_{\infty }}U_{^{\infty }}^{2}{{A}_{ref}}}, \\
		\end{aligned}
	\end{equation}
	where $L$ and $D$ are the lift and drag forces, respectively. The hybrid method demonstrates excellent agreement with benchmark results, showing maximum relative errors of just $2.93\%$ for the lift coefficient and $1.72\%$ for the drag coefficient. These results confirm both the accuracy of the hybrid method and numberical capability to simulate complex three-dimensional flow fields.
    	
	\begin{table}
        \centering
        \caption{\label{X-38 coeff} Comparisons of X-38 like coefficients with Mach number 8}
        \begin{threeparttable}
            \begin{tabular}{p{60pt}<{\centering}|p{80pt}<{\centering}|p{60pt}<{\centering}|p{70pt}<{\centering}|p{60pt}<{\centering}|p{70pt}<{\centering}}
                \hline
                \hline
                Coefficients      &Hybrid method &UGKS &Relative error &RariHV &Relative error   \\
                \hline
                $C_{L}$   & 0.19865     & 0.193    & $2.93 \%$ 
                & 0.194   & $2.4 \%$ \\
                \hline
                $C_{d}$   & 0.29498       & 0.29    & $1.72 \%$
                & 0.295    & $-0.01 \%$ \\
                \hline
                \hline
            \end{tabular}
        \end{threeparttable}
    \end{table}

    \section{Conclusions}\label{sec:Conclusion}
	In this paper, a novel hybrid particle method is proposed by coupling SP-Shakhov method and DSMC method through a temporal scale-based strategy. The distinguishing feature of this approach is its algorithmic coupling framework within a single computational mesh, where near-continuum behavior is modeled by the SP-Shakhov method, and strongly non-equilibrium effects are captured by DSMC. By decomposing the collision operator into two sub-steps according to the ratio of the local relaxation time to the numerical observation time step, the hybrid method enables a physically consistent and dynamically adaptive allocation between SP and DSMC contributions.

	The proposed method preserves the accuracy of DSMC in rarefied regions while enabling the use of significantly coarser grid resolutions (e.g., $\Delta x = 5\lambda$ in this study), as permitted by the SP-Shakhov method in near-continuum regimes. The accuracy and robustness of the proposed method are validated through a series of benchmark cases spanning a wide range of flow regimes, including the one-dimensional Sod shock tube, two-dimensional hypersonic flow over a cylinder and jet expansion into vacuum, and three-dimensional hypersonic flows over a sphere and an X-38 like vehicle. In all cases, the results show excellent agreement with reference solutions from the UGKS, pure DSMC, and SP-Shakhov method.

	In contrast to conventional domain decomposition strategies that explicitly partition the computational domain into continuum and rarefied regions, the present approach achieves an equivalent separation implicitly by leveraging local temporal scales within a unified mesh framework. In strongly non-equilibrium regions, such as pre-shock and near-wall ones, DSMC contributions naturally dominate, while in far-field regions with increasing grid sizes, the DSMC weight decreases rapidly. A sharp drop in DSMC contribution is also observed behind bow shocks due to flow compression. However, complex geometries may lead to artificial mesh refinement in equilibrium regions, result in undesirably large DSMC weights. To address this issue, future developments will explore adaptive Cartesian mesh refinement techniques to maintain physically consistent spatial scales. Furthermore, replacing the SP method with the USP method will be considered to further enhance the method's applicability to continuum-rarefied multiscale flows.

	\section*{Acknowledgements}
	The authors thank Dr. Guang Zhao for providing the computational mesh of X-38 like vehicle. Thanks to Mr. Rui Zhang and Mr. Jianfeng Chen for providing the UGKS and CDUGKS reference results. This work was financially supported by the National Natural Science Foundation of China (Grants 12172301), and the Program of Introducing Talents of Discipline to Universities (111 Project of China, Grant B17037).

    \section*{DATA AVAILABILITY}
    The data that support the findings of this study are available from the corresponding author upon reasonable request.
	
	\clearpage

    \clearpage


%
%

%


\bibliography{Reference}

\end{document}